\documentclass[aps,prb,showpacs,twocolumn,floats]{revtex4}
\usepackage{amssymb}
\usepackage{amsbsy}
\usepackage{amsmath}
\usepackage{graphicx}
\usepackage{amsmath}
\usepackage{times}
\usepackage{color}
\usepackage{subfigure}
\usepackage{setspace}
\usepackage{bm}% bold math
\newcommand\bea{\begin{eqnarray}}
\newcommand\eea{\end{eqnarray}}
\newcommand\beq{\begin{equation}}
\newcommand\eeq{\end{equation}}
\newcommand\bib{\bibitem}

\newcommand{\noi}{\noindent}
\newcommand{\non}{\nonumber}

\newcommand{\de}{\delta}
\newcommand{\si}{\sigma}

\newcommand{\om}{\omega}
\newcommand{\da}{\dagger}
\newcommand{\la}{\langle}
\newcommand{\ra}{\rangle}
\newcommand{\vk}{{\vec k}}
\newcommand{\vn}{{\vec n}}
\newcommand{\vcr}{\vec r}

\begin{document}

\title{Signatures and conditions for phase band crossings in periodically
driven integrable systems}

\author{Bhaskar Mukherjee$^1$, Arnab Sen$^1$, Diptiman Sen$^2$, and
K. Sengupta$^1$}

\affiliation{$^1$ Theoretical Physics Department, Indian Association for the
Cultivation of Science, Jadavpur, Kolkata 700 032, India \\
$^2$ Centre for High Energy Physics, Indian Institute of Science,
Bengaluru 560 012, India}

\date{\today}

\begin{abstract}

We present generic conditions for phase band crossings for a class
of periodically driven integrable systems represented by free
fermionic models subjected to arbitrary periodic drive protocols
characterized by a frequency $\om_D$. These models provide a
representation for the Ising and $XY$ models in $d=1$, the Kitaev
model in $d=2$, several kinds of superconductors, and Dirac fermions
in graphene and atop topological insulator surfaces. Our results
demonstrate that the presence of a critical point/region in the
system Hamiltonian (which is traversed at a finite rate during the
dynamics) may change the conditions for phase band crossings that
occur at the critical modes. We also show that for $d>1$, phase band
crossings leave their imprint on the equal-time off-diagonal
fermionic correlation functions of these models; the Fourier
transforms of such correlation functions, $F_{\vec k_0}( \omega_0)$,
have maxima and minima at specific frequencies which can be directly
related to $\om_D$ and the time at which the phase bands cross at
$\vk = \vk_0$. We discuss the significance of our results in
the contexts of generic Hamiltonians with $N>2$ phase bands and the
underlying symmetry of the driven Hamiltonian.

\end{abstract}

\pacs{73.43.Nq, 05.70.Jk, 64.60.Ht, 75.10.Jm}

\maketitle

\section{Introduction}
\label{intro}

Non-equilibrium dynamics of closed quantum systems has been a
subject of intense theoretical and experimental research in recent
years \cite{pol1,dziar1,dutta1,marcos1}. Such systems are known to
show several interesting features which have no analogs in their
equilibrium counterparts. Some such phenomena include Kibble-Zurek
scaling of defect density upon passage through a quantum critical
point \cite{kz1,pol2,ks1,pol3,sondhi0} or a critical (gapless)
region \cite{ds1}. In addition, such drives may lead to dynamic
transitions which cannot be characterized by any local order
parameter \cite{mukh1,pol4,dytr1,dytr2,dutta2} but manifest
themselves in the vanishing of the Loschmidt overlap $F(t) = \langle
\psi_i| \exp[-i H_f t] |\psi_i\rangle$, where $|\psi_i\rangle$ is
the initial system wave function (often chosen to be the ground
state of $H_i$) and $H_f$ is the final Hamiltonian following a
quench of a Hamiltonian parameter. Such dynamical phase transitions
can be defined in terms of non-analyticities (also known as Fischer
zeroes) of the dynamical free energy of the system $f(z) = - \lim_{L
\to \infty} \ln(F(z))/L^d$, where $z$ is obtained by analytic
continuation of time $t$ in the complex plane. Finally, quantum
quenches may lead to novel properties of the work distribution of
quantum systems following the quench which are qualitatively
different from their equilibrium counterparts \cite{silva1,silva2}.

Out of the drive protocols studied theoretically and experimentally
so far, periodic drives are found to lead to a gamut of interesting
phenomena which do not have counterparts in aperiodically driven
systems. These include dynamics induced freezing where the state of
the system, after several or single cycle(s) of the drive, has close
to unity overlap with its initial state; such a phenomenon can be
related to Stuckelberg interference between quantum states of the
driven system \cite{arnab1,pekker1,uma1}. In addition, we may use
periodic drives to obtain novel steady states in many-body localized
systems where a fast drive may lead to delocalization while a slow
drive keeps the system localized \cite{arnab2}. Moreover, it was
shown that periodically driven interacting systems may lead to
stable out-of-equilibrium phases in the presence of disorder
\cite{roderich1}; such phases are, similar to their equilibrium
counterparts, amenable to symmetry based classification
\cite{sondhi1}. Furthermore, the work distribution of periodically
driven system shows an oscillatory behavior with the drive
frequency; such a behavior constitutes an example of a quantum
interference effect shaping the behavior of a thermodynamic quantity
in a closed quantum system \cite{arnab3}. Finally, periodically
driven integrable systems are known to undergo a separate class of
dynamical phase transitions; these transitions, in contrast to the
ones discussed for aperiodically driven systems, leave their mark
through a change in the convergence of local correlation functions
to their steady state values; they can be shown to be a consequence
of a change in topology of the Floquet spectrum of the driven system
as a function of the drive frequency \cite{asen1}.

Apart from the effects mentioned earlier, another widely studied
phenomenon that occurs in periodically driven clean quantum systems
involves the generation of topological phases characterized by
non-trivial edge modes even when the starting ground state of the
corresponding equilibrium Hamiltonian is topologically trivial
\cite{kita1,lind1,jiang,gu,kita2,lind2,
morell,trif,russo,basti1,liu,tong,cayssol,rudner,basti2,tomka,gomez,dora,
katan,kundu,basti3,schmidt,reynoso,wu,perez,thakurathi1,reichl,thakurathi2}.
Such systems have been treated both analytically and numerically
demonstrating the appearance of edge modes after a drive through one
or more time periods signifying that the system has entered a
topological phase. Recently, however, a more complete understanding
of generation of edge states due to periodic drives in clean systems
has been put forth in Ref.\ \onlinecite{rudner1} in terms of the
properties of the time evolution operator $U(t,0) \equiv U(t)$ given
by
\begin{eqnarray} U(t) &=& {\mathcal T}_t e^{- (i/\hbar) \int_0^t dt' H(t')},
\quad 0\le t \le T, \label{evop} \end{eqnarray}
where $H(t)$ denotes the periodically driven Hamiltonian of the system,
${\mathcal T}_t$ denotes time-ordering, we have chosen the initial time
$t_i=0$ without loss of generality, and here and in the rest of the work we
shall denote $T= 2\pi/\om_D$ to be the drive period, where $\om_D$ is the
drive frequency. It was pointed out in Refs.\ \onlinecite{rudner1} and
\onlinecite{rudner2} that the knowledge of $U(T)
= \exp[-i H_F T/\hbar]$, or equivalently the Floquet Hamiltonian
$H_F$, is insufficient for describing the topological properties of
the system. Such properties can be understood instead by tracking
the crossings of the phase bands $\phi_n(\vk, t)$ which are defined
using the expression of $U_{\vk}(t)$ for $t \le T$ as
\begin{eqnarray} U_{\vk}(t) = \sum_{n=1}^{n_{\rm max}} P_n(\vk; t) e^{i
\phi_n(\vk; t)}. \end{eqnarray} Here we have assumed that the
crystal momentum $\vk$ is a good quantum number, $n$ is the band
index with $n_{\rm max}$ bands for each ${\vk}$, $\lambda_n(\vec
k, t) = \exp[i \phi_n(\vk,t)]$ are eigenvalues of $U_{\vec
k}(t)$, and $P_n(\vk t)$ projects $U_{\vk}(t)$ to its $n^{\rm
th}$ eigenstate. We note that $U_{\vk}(0)=1$ indicates that
$\phi_n(\vk, 0) = 2 \pi m$, where $ m \in Z$. Thus the phase
bands may be represented either in the repeated zone scheme or the
reduced zone scheme where the $(n_{\rm max} +1)^{\rm th}$ band is
identified with the $n=1$ band. In what follows, we shall adopt the
latter representation.

It was shown in Ref. \onlinecite{rudner1} that the topological
properties of such periodic driven systems may be understood in
terms of phase band crossings. As argued in Ref.\
\onlinecite{rudner1}, phase band crossings are topologically
significant only if they occur between the first and the top bands
of the reduced Brillouin zone; all other crossings can be gauged
away by simple deformations of the drive protocol. Each such
topologically non-trivial crossing is associated with a finite
topological charge $q_i$; the number of edge modes which result from
such a crossing can be directly related to $q_i$. For example in
$d=2$, where each phase band can be represented by a Chern number
$C_n$, the number of chiral edge states within the $m^{\rm th}$ bulk
Floquet band is given by $n_{\rm edge}(m) = \sum_{n=1, M} C_n -
\sum_i q_i$. Further, it was shown in several earlier works
\cite{jiang,rudner1,carp1} that the presence of particle-hole and
time-reversal symmetries may lead to further restrictions on such
crossings; for example, in the presence of particle-hole symmetry
and for one-dimensional (1D) driven Hamiltonians, the phase band
crossings can occur only at $k_0=0$ or $\pi/a$, where $a$ is the
lattice spacing of the model (which we will subsequently set equal
to $1$, unless mentioned otherwise). The number of edge modes in
these systems are completely determined by the parity of the number
of such crossings at $k_0=0$ and $\pi$ \cite{rudner1,comment1}.
However, the earlier works on phase band crossing did not
systematically study the role of the drive protocol; further the
conditions for such crossings has not been methodically investigated
in terms of the parameters of the driven Hamiltonian beyond a few
simple protocols and toy models \cite{kita1,lind1,jiang,zhao}. In
this work, we aim to fill up this gap.

To this end, we study a class of periodically driven integrable
models whose Hamiltonian can be represented by free fermions in
$d$-dimensions:
\begin{eqnarray}
H(t)= \sum_{\vk} \psi_{\vk}^{\dagger} H_{\vk}(t) \psi_{\vec
k}, \label{fermham}
\end{eqnarray}
where $\psi_{\vk}= (c_{\vk}, c_{-\vk}^{\dagger})^T$ is the
two-component fermionic field, $c_{\vk}$ are the annihilation
operators for the fermions, and $H_k(t)$ is given by
\begin{eqnarray} H_{\vk} = (g(t) - b_{\vk}) \tau^z + (\Delta_{\vk}
\tau^+ + \rm{H.c.} ), \label{fermhamden} \end{eqnarray} where $g(t)$
is a periodic function of time characterized by a frequency
$\om_D$, and $\Delta_{\vk}$ can be an arbitrary function of
momenta. We note that this kind of Hamiltonian represents a wide
class of spin and fermionic integrable models such as the Ising and
$XY$ models in $d=1$ \cite{subir1}, the Kitaev model in $d=2$
\cite{kit1,nussinov1}, triplet and singlet superconductors in $d>1$,
and Dirac fermions in graphene and atop topological insulator
surfaces \cite{graphene1,topo1}. In what follows, we shall obtain
our results by analyzing fermionic systems given by Eq.\
\eqref{fermhamden} and point out the relevance of these results in
the context of specific models in appropriate places.

The main results that we obtain from such an analysis are the
following. First, we obtain an expression for the phase bands
corresponding to Hamiltonians given by Eq.\ \eqref{fermhamden}
within the adiabatic-impulse approximation \cite{nori1,kanu1,child1}
for arbitrary drive protocols. Using these expressions and other
general arguments, we chart out the most general conditions that
need to be satisfied for these phase bands to cross. The conditions
that we obtain conform to those obtained earlier for particle-hole
symmetric Hamiltonians \cite{rudner1} and for specific drive
protocols \cite{thakurathi1,thakurathi2}; however it constitutes a
more general result which holds for arbitrary periodic drive
protocols and irrespective of the symmetry of the underlying
Hamiltonian. Second, we show that traversing a critical point during
such periodic dynamics may lead to qualitatively different band
crossing conditions, and we discuss its implications for the
properties of the driven system. Third, for $d>1$, we compute the
off-diagonal fermionic correlation function $F_{\vk}(t) = \langle
c_{\vec k}^{\dagger} c_{-\vk}^{\dagger} \rangle$ and show that the
Fourier transform of this correlator will exhibit maxima and minima
at specific frequencies $\omega_0$ for $\vec k= \vec k_0$ at which
the bands cross; we provide an explicit relation between $\om_0$,
$\om_D$ and the band crossing time $t_0$ for several drive
protocols. Thus we show that $F_{\vec k_0}(\om_0)$ carries
information about the phase band crossing time $t_0$. Finally, we
comment on the applicability of our results to general Hamiltonians
with $N>2$ phase bands and present a discussion of the role of
symmetries of the underlying Hamiltonian in the phase band
crossings.

The plan of the rest of the paper is as follows. In Sec.\
\ref{pcross}, we derive explicit expressions for the phase bands
within adiabatic-impulse approximation, obtain the conditions for
their crossings, and point out the role of critical points for such
crossings. This is followed by Sec.\ \ref{corr1}, where we chart out
the behavior of $F_{\vk_0}(\omega_0)$ and discuss the signatures of
phase band crossings which can be inferred from its behavior.
Finally, we discuss the significance of our results for more general
phase band models, point out the role of symmetries for such
crossings, and conclude in Sec.\ \ref{diss}.

\section {Phase band crossings}
\label{pcross}

In this section, we first obtain an expression for the phase bands
corresponding to the Hamiltonian in Eq.~\eqref{fermhamden} within the
adiabatic-impulse approximation for an arbitrary continuous time
protocol in Sec.\ \ref{ad-imp}. This will be followed by an analysis of
the obtained expression for $U_{\vk}(t)$ leading to the phase band
crossing conditions in Sec.\ \ref{pbcr}.

\subsection{Expression for the phase bands}
\label{ad-imp}

To obtain an expression for $U_{\vk}(t)$ for an arbitrary drive
protocol $g(t)$, which is characterized by a frequency $\om_D$, we
use an adiabatic-impulse approximation which has been used
extensively for two-level systems \cite{nori1,kanu1,child1}. We
envisage a drive protocol which starts at $t_i=0$ and continues till
the end of one drive period $t_f= T$. In the rest of this section,
we shall mostly work in the adiabatic basis in which the wave
function at any time $t$ is given by
\begin{eqnarray} |\psi_{\vk}\rangle (t) &=& c_{1 \vk}(t) \left(
\begin{array}{c} u_{0\vk}(t) \\ v_{0\vk}(t) \end{array} \right)
+ c_{2 \vk}(t) \left(
\begin{array}{c} -v_{0\vk}(t) \\ u_{0\vk}(t) \end{array} \right),
\label{adbase} \end{eqnarray}
where $|\psi_{\vk}^g (t)\rangle= (u_{0\vk}(t), v_{0\vk}(t))^T$ and
$-E_{\vk}(t)$ are the instantaneous ground state wave function and energy
which are given by
\begin{eqnarray}
u_{0\vk}(t) &=& -\frac{\Delta_{\vk}}{D_{\vk}(t)}, \quad v_{\vec
k}(t) = \frac{E_{\vk}(t)+g(t)-b_{\vk}}{D_{\vk}(t)}, \non \\
E_{\vk}(t)&=& \sqrt{(g(t)-b_{\vk})^2 + |\Delta_{\vk}|^2},
\label{wav1} \\
D_{\vk}(t) &=& \sqrt{( E_{\vk}(t)+ g(t)-b_{\vk})^2 +
|\Delta_{\vk}|^2}. \non \end{eqnarray}
The corresponding excited state wave function and energies are given
by $|\psi_{\vk}^e (t)\rangle= (-v_{0\vk}(t), u_{0\vec
k}(t))^T$ and $E_{\vk}(t)$. We note here that the adiabatic and
the diabatic bases are connected by the standard transformation
\begin{eqnarray} \left( \begin{array}{c} |\psi_{\vk}^g(t)\rangle \\
|\psi_{\vk}^e (t) \rangle \end{array} \right) &=& \left( \begin{array}{cc}
\mu_{\vk}(t) & \sqrt{1-\mu_{\vk}^2(t)} \\
-\sqrt{1-\mu^2_{\vk}(t)} & \mu_{\vk}(t) \end{array} \right)
\left( \begin{array}{c} |\psi_{\vk}^g\rangle \\
|\psi_{\vk}^e \rangle \end{array} \right) \non \\
\mu_{\vk}(t) &=& u_{0\vk}(t) u_{0\vk} + v_{0\vk}(t) v_{0\vk},
\label{addirel} \end{eqnarray} where $u_{0\vk} \equiv u_{0\vk
}(t=0)$, similar notations have been used for all other quantities,
and in the rest of this section we shall assume the system to be in
its initial ground state at $t=0$. The unitary evolution operator
$U_{\vk}(t)$ relates, by definition, the wave function at time $t$
to the initial wave function and thus satisfies
\begin{eqnarray} |\psi_{\vk} (t)\rangle &=& U_{\vk} (t)
|\psi_{\vk}^g\rangle. \label{evoldef} \end{eqnarray}

\begin{figure}
\includegraphics[width=\linewidth]{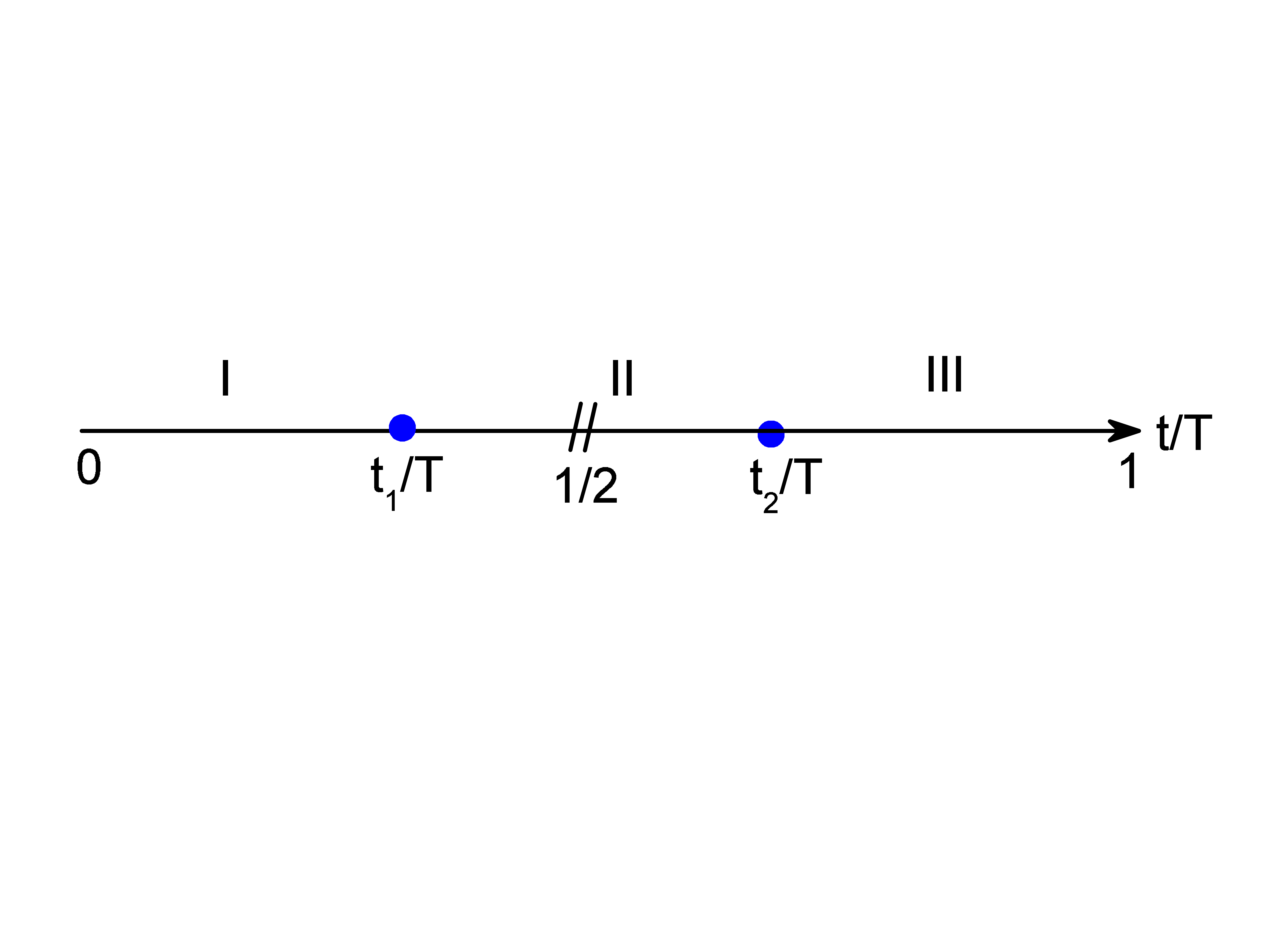}
\caption{Schematic representation of the time-evolution of a
periodically driven system for a drive cycle $T= 2 \pi/\om_D$.
The system reaches the critical points at $t=t_{1 \vk} \equiv
t_1$ and $t_{2 \vk} \equiv t_2$. The half-period $t= T/2$ is
marked with a hash. The adiabatic regimes before and after the first
crossing of critical point are marked as regions I and II respectively,
while that after the second crossing of the critical point is marked
as region III. See text for details.} \label{fig1} \end{figure}

The time evolution of a system described by Eq.\ \eqref{fermhamden}
is sketched in Fig.\ \ref{fig1}. We divide the time evolution into
three distinct regions as shown in Fig.\ \ref{fig1}. Within the
adiabatic-impulse approximation, regions I, II and III are adiabatic
regions. In these regions, the system is sufficiently far way from
the critical points, crossed at times $t_{1 \vk}$ and $t_{2 \vec
k}$, so that the instantaneous energy gap for any $\vk$ satisfies
the Landau criterion: $2 E_{\vk}^2 (t) \gg dE_{\vk}(t)/dt$. It can
be shown that in this regime the system merely gathers kinetic phase
\cite{nori1}
\begin{eqnarray} U'_{\vk} (t_f, t_i) &=& \exp[ - i \xi_{\vk}(t_f,t_i)
\tau^z], \non \\
\xi_{\vk} (t_f,t_i) &=& \int_{t_i}^{t_f} dt E_{\vk}(t) \non \\
\left( \begin{array}{c} c_{1 \vk}(t_f) \\ c_{2 \vk}(t_f) \end{array}
\right) &=& U'_{\vk}(t_f,t_i) \left( \begin{array}{c}
c_{1 \vk}(t_i) \\ c_{2 \vk}(t_i) \end{array} \right). \label{evo1}
\end{eqnarray}

In the impulse region, the adiabaticity condition given by the
Landau criteria breaks down. For slow enough drives, this happens
around the critical point and for momentum modes sufficiently close
to the critical mode. The key idea of the adiabatic-impulse
approximation is to approximate the impulse region to be an
infinitesimally small region around the critical point; such an
approximation holds good for large amplitude and low frequency
drives \cite{nori1,kanu1,child1}. The impulse region is typically
reached twice during a drive cycle for any momenta $\vk$, at $t=t_{1
\vk}$ and $t_{2 \vk}= 2\pi/\om_D -t_{1 \vk}$, where
$g(t_{1(2)\vk})=b_{\vk}$. Around $t=t_{1(2) \vk}$, we can linearize
the diagonal terms of the Hamiltonian (Eq.\ \eqref{fermhamden}) and
obtain
\begin{eqnarray} H_{\vk}^{\rm imp} &=& v_{\vk} (t-t_{1(2) \vk})
\tau^z + (\Delta_{\vk} \tau^+ + {\rm H.c.}), \label{impham} \end{eqnarray}
where $v_{\vk}= \partial g(t_{1(2)\vk})/\partial t$.
\cite{comment2} The probability of excitation between the ground
and excited states at each $\vk$ can be read off from Eq.\
\eqref{impham} as \cite{nori1}
\begin{eqnarray} p_{\vk} &=& \exp[-2 \pi \delta_{\vk}], \quad
\delta_{\vk} = |\Delta_{\vk}|^2/|2v_{\vk}|. \label{probex}
\end{eqnarray}
We note that $p_{\vk}=1$ for an unavoided level crossing which
happens for the critical mode (where $\Delta_{\vk}=0$ and $g(t) -
b_{\vk}$ crosses zero), while it is zero if $\Delta_{\vk}=0$ but
$g(t) \ne b_{\vk}$ at any point of time during the drive. It was
shown in Ref.\ \onlinecite{kanu1} that within this approximations we
can define a transfer matrix
\begin{eqnarray} S_{\vk} &=& \left( \begin {array} {cc} \sqrt{1-p_{\vk}}
e^{-i \tilde \phi_{s\vk}} & - \sqrt{p_{\vk}} \\ \sqrt{p_{\vk}} &
\sqrt{1-p_{\vk}}e^{i \tilde \phi_{s\vk}} \end {array} \right),
\label{evo4} \\
\phi_{s \vk} &=& \frac{\pi}{4} + \delta_{\vk} \left( \ln
\delta_{\vk} -1 \right) + {\rm Arg} ~\Gamma (1 - i \delta_{\vk} ),
\label{smatrix1} \end{eqnarray} where $\phi_{s \vk}$ is the Stoke's
phase and $\tilde \phi_{s\vec k} = \phi_{s \vec k}-\pi$. The change
of wave functions across the first transition point can be obtained
using the transfer matrix $S_{\vk}$ as
\begin{eqnarray} \left( \begin{array}{c} c_{1 \vk}(t_{1 \vk} +
\epsilon) \\
c_{2 \vk}(t_{1 \vk} + \epsilon) \end{array} \right) &=& S_{\vk}
\left( \begin{array}{c} c_{1 \vk}(t_{1 \vk} -\epsilon) \\
c_{2 \vk}(t_{1 \vk} - \epsilon) \end{array} \right), \label{evo2}
\end{eqnarray}
where $\epsilon>0$ is infinitesimally small. An analogous condition
with $S$ replaced by $S^T$ (where the superscript $T$ denotes
transpose) holds for the second transition point.

Having obtained these relations, we now obtain explicit analytical
expressions for the evolution operator $U_{\vk}(t)$ in each of the
three regions shown in Fig.\ \ref{fig1}. In region I, the system
starts from the ground state so that $c_{1\vk}(0)=1$ and $c_{2
\vk} = 0$, Before crossing the critical point for the first time, the
dynamics is purely adiabatic leading to (using Eqs.~ \eqref{adbase} and
\eqref{evo1})
\begin{eqnarray} c_{1\vk}^I(t) &=& \exp[-i\xi_{\vk} (t,0)], \quad
c_{2 \vk}^I (t)=0. \label{creg1} \end{eqnarray}
Using Eqs.\ \eqref{adbase}, \eqref{addirel} and \eqref{creg1}, we thus obtain
\begin{eqnarray} |\psi_{\vk}(t)\rangle &=& e^{-i\xi_{\vk} (t,0)} \left(
\mu_{\vk}(t) |\psi_{\vk}^g\rangle + \sqrt{1-\mu_{\vk}(t)^2}
|\psi^e_{\vk}\rangle \right). \label{wavreg12} \non \\
\end{eqnarray}
Using Eqs.~\eqref{evoldef} and \eqref{wavreg12} and the fact that
$U_{\vk}(t)$ is a unitary matrix, we obtain for region I
\begin{eqnarray} U_{\vk}^{I}(t) &=& \left( \begin{array}{cc} \mu_{\vk}(t)
e^{-i\xi_{\vk} (t,0)} & -\sqrt{1-\mu_{\vk}^2(t)} e^{i\xi_{\vk}
(t,0)} \\
\sqrt{1-\mu_{\vk}^2(t)} e^{-i\xi_{\vk} (t,0)} & \mu_{\vk}(t)
e^{i\xi_{\vk} (t,0)} \end{array} \right). \label{uexpreg1} \non \\
\end{eqnarray}
The eigenvalues $\lambda_{\pm \vk}^{I}(t)$ of $U_{\vk}^{I}(t)$
provide an expression for the phase bands in region I and are given by
\begin{eqnarray} \lambda_{\pm \vk}^{I}(t) &=& e^{\pm i \phi_{\vk}^{I}
(t)}, \non \\
\cos[\phi_{\vk}^I(t)] &=& \mu_{\vk}(t) \cos(\xi_{\vk}(t,0)).
\label{eigenreg1} \end{eqnarray}

An exactly similar method can be used to compute the evolution
operators in regions II and III. For example, in region II, using
Eqs.\ \eqref{adbase}, \eqref{evo1}, \eqref{evo4}, and
\eqref{smatrix1}, we obtain
\begin{eqnarray} \left( \begin{array}{c} c_{1\vk}^{II}(t) \\
c_{2\vk}^{II} (t) \end{array} \right) &=& U'_{\vk}(t,t_{1\vec
k}) S_{\vk} U'_{\vk}(t_{1 \vk},0) \left( \begin{array}{c} 1\\
0 \end{array} \right), \label{ceqreg2a} \end{eqnarray}
which leads to
\begin{eqnarray} c_{1\vk}^{II}(t) &=& \sqrt{1-p_{\vk}} e^{-i
\zeta_{1 \vk}^{II}(t)}, \quad c_{2\vk}^{II}(t) = \sqrt{p_{\vk}} e^{i
\zeta_{2 \vk}^{II}(t)}, \non \\
\zeta_{1 \vk}^{II}(t) &=& \tilde \phi_{s\vk} + \xi_{\vk}(t,0), \non \\
\zeta_{2 \vk}^{II}(t) &=& \xi_{\vk}(t_{1\vk},0)-\xi_{\vk}
(t,t_{1\vk}). \label{ceqreg2b} \end{eqnarray}
Using Eqs.~\eqref{addirel}, \eqref{evoldef} and \eqref{ceqreg2b}, we then
obtain
\begin{eqnarray} (U^{II}_{\vk}(t))_{11} &=& c_{1\vk}^{II}(t)
\mu_{\vk}(t) + c_{2\vk}^{II}(t) \sqrt{1-\mu_{\vk}^2(t)}, \non \\
(U^{II}_{\vk}(t))_{21} &=& c_{2\vk}^{II}(t) \mu_{\vk}(t) -
c_{1\vk}^{II}(t) \sqrt{(1- \mu_{\vk}^2(t)}, \label{uexpreg2} \\
(U^{II}_{\vk}(t))_{22}^{\ast} &=& (U^{II}_{\vk}(t))_{11},
\quad (U^{II}_{\vk}(t))_{12}^{\ast} = -(U^{II}_{\vk}(t))_{21}. \non
\end{eqnarray}
The expression for the phase bands in region II may be obtained by
diagonalizing the evolution matrix $U_{\vk}^{II}(t)$. A
straightforward calculation yields the expressions for the
eigenvalues of $U_{\vk}^{II}(t)$ as $ \lambda_{\pm \vec
k}^{II}(t) = \exp[\pm i \phi_{\vk}^{II}(t)]$, where
\begin{eqnarray} \cos[\phi_{\vk}^{II}(t)] &=& \mu_{\vk}(t)
\sqrt{1-p_{\vk}} \cos(\zeta_{1\vk}^{II}(t)) \non \\
&& + \sqrt{p_{\vk}(1-\mu_{\vk}^2(t))} \cos(\zeta^{II}_{2\vk}(t)).
\label{eigenreg2} \end{eqnarray}

A similar calculation can be carried out in region III. Here one
finds that
\begin{eqnarray} \left( \begin{array}{c} c_{1\vk}^{III}(t) \\
c_{2\vk}^{III} (t) \end{array} \right) &=& U'_{\vk}(t,t_{2\vec
k}) S^T_{\vk} U'_{\vk}(t_{2 \vk},0) \left( \begin{array}{c} 1\\
0 \end{array} \right), \label{ceqreg3a}
\end{eqnarray}
where $S_{\vk}^T$ is the transpose of $S_{\vk}$ (Eqs.\ \eqref{evo4}
and \eqref{smatrix1}). A straightforward calculation then leads to
\begin{eqnarray} c_{1\vk}^{III}(t) &=& (1-p_{\vk}) e^{-i \zeta_{1 \vec
k}^{III}(t)} + p_{\vk} e^{-i \zeta_{2 \vk}^{III}(t)}, \non \\
c_{2\vk}^{III}(t) &=& \sqrt{p_{\vk}(1-p_{\vk})} e^{i
\phi_{s \vk}} \left( e^{-i \zeta_{1 \vk}^{III}(t)} - e^{-i
\zeta_{2 \vk}^{III}(t)}
\right), \non \\
\zeta_{1 \vk}^{III}(t) &=& 2 \tilde \phi_{s\vk} + \xi_{\vk}(t,0), \non \\
\zeta_{2 \vk}^{III}(t) &=& \xi_{\vk}(t_{1\vk},0)-\xi_{\vk}(t_{2
\vk},t_{1\vk}) + \xi_{\vk}(t,t_{2\vk}). \label{ceqreg3b}
\end{eqnarray}
Using Eqs.~\eqref{addirel}, \eqref{evoldef} and \eqref{ceqreg2b}, we
obtain $U_{\vk}^{III}(t)$ in an analogous manner. The
expressions for the phase bands in region III are then obtained by
diagonalizing $U_{\vk}^{III}$ and yield $ \lambda_{\pm \vec
k}^{III}(t) = \exp[\pm i \phi_{\vk}^{III}(t)]$, where
\begin{eqnarray} && \cos[\phi_{\vk}^{III}(t)] = \mu_{\vk}(t) \left(
(1-p_{\vk})\cos(\zeta_{1\vk}^{III}(t)) \right. \non \\
&& + \left. p_{\vk} \cos(\zeta_{2\vk}^{III}(t))\right) +
\sqrt{p_{\vk}(1-p_{\vk})(1-\mu_{\vk}^2(t))} \non \\
&& \times \left( \cos(\zeta^{III}_{1\vk}(t) - \tilde \phi_{s\vk})-
\cos(\zeta^{III}_{2\vk}(t) - \tilde \phi_{s\vk}) \right).
\label{eigenreg3}
\end{eqnarray}
Eqs.\ \eqref{eigenreg1}, \eqref{eigenreg2}, and \eqref{eigenreg3}
constitute the central results of this section. These equations
provide us with analytic expressions for the phase bands for
integrable models studied for an arbitrary continuous time protocol.
In what follows, we shall analyze these expressions to obtain
conditions for phase band crossings for these models.

\begin{figure}
\includegraphics[width=\linewidth]{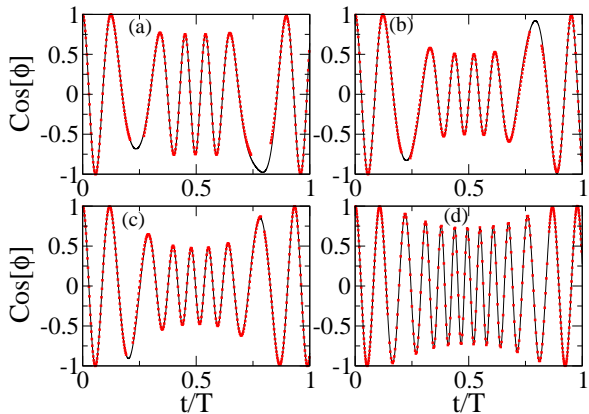}
\caption{A plot of $\cos(\phi_k(t))$ as a function of $t/T$ for (a)
$k=\pi/20$, (b) $k=\pi/10$, (c) $k= \pi/6$, and (d) $k= \pi/3$ for
the 1D Ising model in a transverse field for the drive protocol
$h(t)= h_0 + h_1 \cos(\om_D t)$ with $h_0=1.1$, $h_1=-1$, and
$\om_D=0.1$ with all energy scales measured in units of $J$, and
$\hbar$ is set equal to unity. The black solid line correspond to
exact numerical solution of Eqs.\ \eqref{fermsch}, while the red
dotted lines correspond to those obtained using adiabatic-impulse
approximation (Eqs.\ \eqref{eigenreg1}, \eqref{eigenreg2}, and
\eqref{eigenreg3}). The match between the two deteriorates with
increasing $\omega_D$ except at $k=0, \pi$ where the approximation
yields exact results.} \label{fig2} \end{figure}

\begin{figure}
\includegraphics[width=0.47\linewidth]{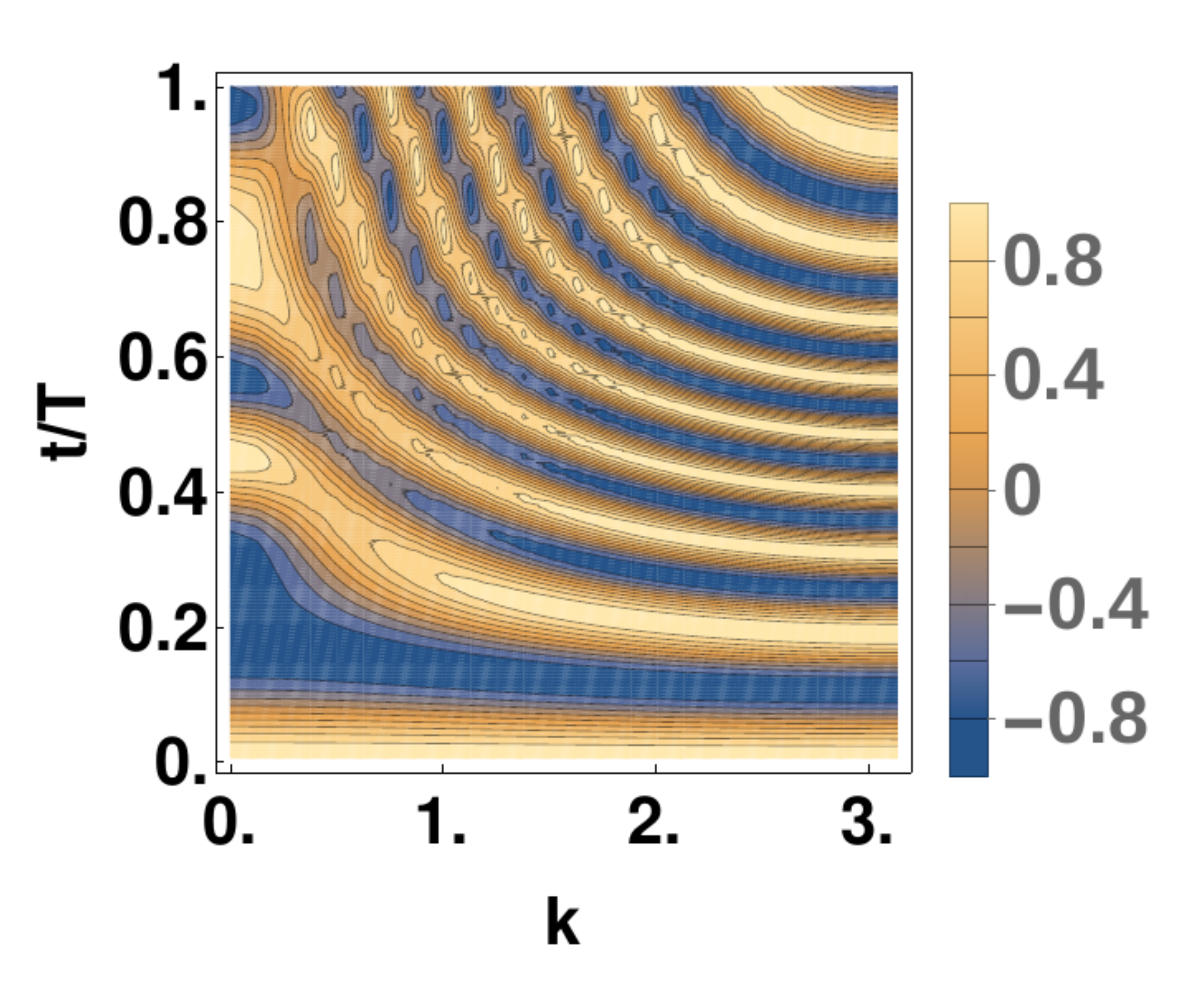}
\includegraphics[height= 3.4 cm, width=0.47\linewidth]{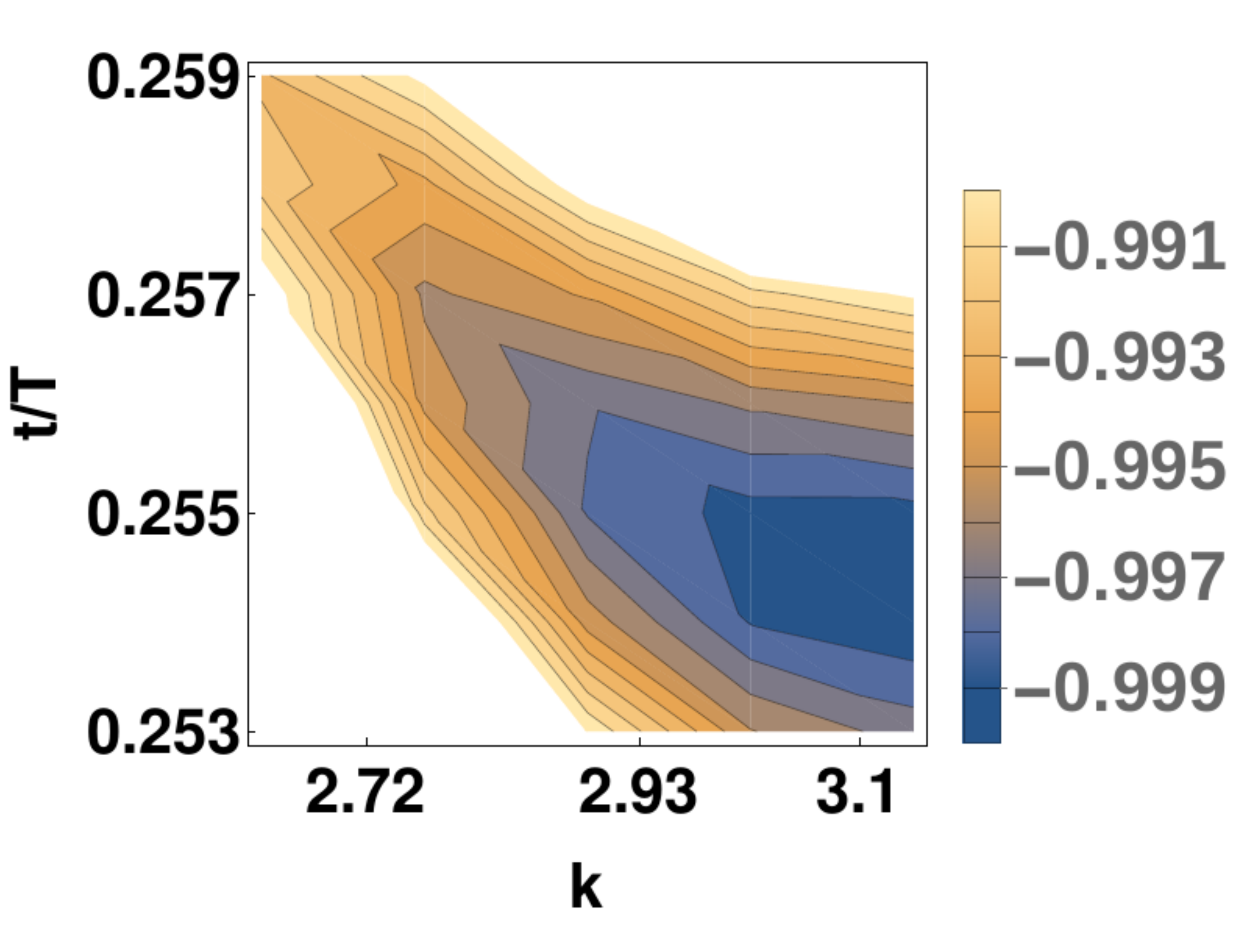}
\caption{Left Panel: Plot of $\cos(\phi_{k}(t))$ as a function of $k$ and
$t/T$ for the ID Ising model with $\omega_D=0.25$. All other parameters are
the same as in Fig.\ \ref{fig2}. Right Panel: A close-up of $\cos \phi_{k}(t)$
showing a phase-band crossing at $k=\pi$ and $t/T \simeq 0.245$. All
parameters are the same as those in the left panel.} \label{fig3} \end{figure}

Before ending this section, we provide a comparison between the
exact numerical values of the phase bands with those obtained by our
method for the $d=1$ Ising model in a transverse field. As is
well-known, Eq.\ \eqref{fermhamden} provides a fermionic
representation of the transverse field Ising model with $b_{k}= \cos
k$, $\Delta_{k} = \sin k$ and $g(t)=h(t)$, where we have set the
nearest-neighbor coupling $J$ between the spins and the lattice
spacing $a$ to unity \cite{subir1}. The critical point for this
system is located at $h=1$. In what follows, we choose $h(t)=h_0 +
h_1 \cos(\om_D t)$ with $h_0=1.1$ and $h_1=-1$ so that the system
traverses the critical point twice during the dynamics. To obtain
the phase bands, we note that the Schr\"odinger equation
corresponding to the fermionic Hamiltonian in Eq.\
\eqref{fermhamden} is given by
\begin{eqnarray} i \partial_t u_k(t) &=& (h(t)-\cos k) u_k + i \sin k \, v_k,
\non \\
i \partial_t v_k (t) &=& -(h(t)-\cos k) v_k - i \sin k \,u_k. \label{fermsch}
\end{eqnarray}
We solve these equations numerically with the initial condition
$(u_k(0),v_k(0))=(u_{0 k}, v_{0 k})$ and obtain the wave function
$(u_k(t),v_k(t))$ at any time $t \le T$. The unitary evolution
operator $U_k(t)$ is then obtained using Eq.\ \eqref{evoldef} from
the initial and final wave functions. Finally, we diagonalize
$U_k(t)$ to obtain its eigenvalues and hence the phase bands
$\lambda_{k \pm}(t) = \exp[\pm i \phi_k(t)]$. A plot of
$\cos(\phi_k(t))$ obtained in this manner is compared to their
adiabatic-impulse counterparts for several representative values of
$k$ and for $\omega_D=0.1$ as shown in Fig.\ \ref{fig2}; the figure
shows a near exact match between the two for all $k$. This feature
is expected since the adiabatic-impulse approximation becomes
accurate for small $\omega_D$ for any $k$; we note here that it is
exact at $k=0, \pi$ for all $\omega_D$. Thus our analytical approach
provides a decent approximation to the phase bands for all $k$ at
low drive frequencies. The structure of these phase bands as a
function of $k$ and $t/T$ is shown in Fig.\ \ref{fig3} for
representative values of $\om_D$ and $h(t)$; we note that these
bands never reach the values 0 or $\pm \pi$ unless $k=0, \, \pi$. We
shall analyze this fact in detail in the next section.

\subsection{Phase band crossing conditions}
\label{pbcr}

The phase bands $\lambda_{ \vk \pm}^{a}(t) = \exp[\pm i
\phi^a_{\vk} (t)]$ (where $a=$ I, II or III) whose expressions
were obtained in Sec.\ \ref{ad-imp}, cross when $\phi_{\vec
k}^{a}(t) = n \pi$ for any integer $n$. We note that all such
crossings for the integrable models that we study constitute
examples of zone-edge singularities \cite{rudner1} and are therefore
topologically significant. To understand when such crossings can
happen, we first note that $\mu_{\vk}(t) \le 1$ since it denotes the
overlap between ground state wave functions at different times. This
property of $\mu_{\vk}(t)$ ensures that the right hand sides of Eqs.\
\eqref{eigenreg1}, \eqref{eigenreg2}, and \eqref{eigenreg3} can becomes
unity only when $\mu_{\vk}(t)= 1 \, {\rm or} \, 0$. This, in turn,
can occur only for momenta $\vk = \vk_0$ for which $\Delta_{\vk_0}
=0$; thus phase band crossings only occur at these momenta.

This condition for phase band crossings happens to be a general
result which may also be understood from simpler intuitive
arguments. We present two such equivalent arguments here. The first
of these involves a geometrical construction which invokes the
concept of the Bloch sphere. Since the generators of $U_{\vk}(t)$
belong to the SU(2) algebra, the trajectory of any wave function
under its action can be considered as a trajectory on the Bloch
sphere characterized by a fixed momentum $\vk$. Thus the requirement
for a phase band crossing at $U_{\vk_0}(t_0)$ amounts to the
condition $|\psi_{\vk_0}(0)\rangle = |\psi_{\vec k_0}(t_0)\rangle$,
{\it i.e.}, the trajectory of the wave function for $\vk = \vk_0$
must cross itself at $t=t_0$ under the action of $U_{\vk_0}(t_0)$.
Since the eigenvalues of $U_{\vec k}(t)$ are independent of the
initial wave function $|\psi_{\vec k}(0)\rangle$, this condition
requires that the trajectory on the Bloch sphere, no matter where it
starts, must pass through itself at $t=t_0$ during its evolution.
This condition can be generically satisfied if that trajectory is
generated by rotation around a single axis {\it at all times}. Thus
such crossings can only occur if $\Delta_{\vk}=0$.

The second, equivalent, argument showing that $\Delta_{\vk}=0$
constitutes a necessary condition for phase band crossings is as
follows. A Trotter decomposition of Eq.\ \eqref{evop} enables us to
write $U_{\vk}(t)$ at the time $t_0$ when the phase band crosses at
any given $\vk$ as
\begin{eqnarray} U_{\vk}(t_0,0) &=& \lim_{N \to \infty} \prod_{j=0, N-1}
U_{\vec k} (t_{j+1},t_j) \label{trot1} \\
&=& \lim_{N \to \infty} \prod_{j= \ell+1, N-1} U_{\vk}
(t_{j+1},t_{j}) \prod_{j=0, \ell } U_{\vk} (t_{j+1},t_j), \non
\end{eqnarray} where $t_j-t_{j-1} = \Delta t= t_0/N$, $t_0=t_{N}$,
and in the second line we have organized the product into two groups
for which $t_j \le t_{\ell}$ and $t_j > t_{\ell}$. Note that the
choice of $ t_{\ell}$ is completely arbitrary. Now since the product
of these evolution matrices in each of the two groups must also be a
SU(2) rotation matrix, we can write
\begin{eqnarray} U_{\vk}(t_0,0) &=& U_{2 \vec k}(t_0,t_{\ell+1})
U_{1 \vk}(t_{\ell+1},0) \non \\
&=& e^{-i(\vec \sigma \cdot \vn_{2\vk})\phi_{2 \vk}} e^{-i(\vec
\sigma \cdot \vn_{1 \vk})\phi_{1 \vk}}, \label{trot2} \end{eqnarray}
where $\vn_{1(2) \vk}$ are unit vectors and $\phi_{1(2) \vk}$ are
rotation angles. All of these parameters depend, in general, on the
choice of $t_{\ell}$, the details of $H_{\vk}(t)$, and the drive
protocol. We do not attempt to compute them here; instead, we merely
observe that in order to get $U_{\vk}(t_0)=I$ {\it independent of
the choice of $t_{\ell}$}, we clearly require $\vn_{2 \vk} = \vn_{1
\vk}$. This is most easily seen by choosing $n_{1 \vec k}= \hat z$
which can be done without loss of generality by choosing suitable
axes, and then checking that the eigenvalues of $U_{\vec k}$ can
never be unity if $n_{2 \vec k} \ne \hat z$. Next we note that the
condition $n_{2 \vec k}= n_{1 \vec k}$ can only be satisfied for
arbitrary $t_{\ell}$ if $U_{\vk}(t_i)$ commutes at all $t_i$, {\it
i.e.}, if the rotation axis of $U_{\vk}(t_{j+1},t_j)$ is the same
for all $t_j$. In the context of the Hamiltonian $H_{\vk}(t)$ given
in Eq.\ \eqref{fermhamden}, this is only possible if
$\Delta_{\vk}=0$. In a more general context, the Hamiltonian of the
system can be written as
\begin{eqnarray} H_{\vk}(t) &=& \sum_{i=1}^3 \tau_i f_{i \vk}(t),
\label{genham} \end{eqnarray} where $f_{i \vec k}(t)$ are parameter
functions and the Pauli matrices $\tau_i$ are the generators. The
condition for the phase band crossings requires that any two of the
three parameter functions $f_{i \vk}(t)$ vanishes at all times. The
third parameter function then determines $t_0$, and we discuss this
issue in detail below. Finally, we note that the above mentioned
arguments indicate that for arbitrary single-rate protocols and for
$d=1$, the phase band crossings can therefore only occur at $k=0$ or
$\pi$. In contrast, for $d>1$ systems such crossings can occur for a
wider range of momenta.

Next, we chart out the condition for phase band crossing at $\vk=
\vk_0$ which yields the band crossing time $t_0$ for each of the
regions shown in Fig.\ \ref{fig1}. In region I, we find from
Eqs.~\eqref{evo1} and \eqref{eigenreg1} that the phase bands will
cross at time $t_0 < t_{1\vk_0}$ provided that
\begin{eqnarray} \xi_{\vk_0}(t_0,0)= n \pi , \quad n \in Z. \label{croscon1}
\end{eqnarray}

To find the condition for phase band crossings in region II, we need
to find $p_{\vk_0}$. Since at $\vk= \vk_0$, $\Delta_{\vec
k_0}=0$, $p_{\vk_0}$ may assume the values $1$ or $0$. The former
occurs at the critical mode where the instantaneous energy levels of
$H$ cross while for the latter they do not cross. In what follows, we
denote the momenta of the critical modes to be $\vk_{0}$ and for
that of the non-critical modes to be $\vk'_0$. Within our
convention, in $d=1$, $k_0=0$ and $k'_0= \pi$. For the critical
modes, the crossing of the instantaneous energy levels ensures that
there is no overlap of the instantaneous ground state wave function
in region II with the initial ground state wave function. This
corresponds to $\mu_{\vk_0}(t)=0$ and leads to $\phi_{\vec
k_0}^{II}(t) = \zeta^{II}_{2\vk_0}(t)$. Thus using Eqs.\
\eqref{ceqreg2b} and \eqref{eigenreg2}, we obtain
\begin{eqnarray} \xi_{\vk_0}(t_{1 \vk_0},0)-\xi_{\vk_0}(t_0,
t_{1 \vk_0}) &=& n \pi. \label{croscon2} \end{eqnarray}
For the non-critical momenta $k'_0$, when the instantaneous energy
bands do not cross even though $\Delta_{k'_0}$ vanishes, $p_{\vec
k'_0}=0$ and $\mu_{\vk'_0}(t)=1$ in region II. Using Eq.\
\eqref{eigenreg2}, we find that in this case $\phi_{\vk'_0}^{II}(t) =
\zeta^{II}_{1\vk'_0}(t)$, and the phase bands cross if Eq.\ \eqref{croscon1}
holds, with $t_{1 \vk_0} \le t_0 \le t_{2 \vk_0}$.

Finally, we obtain the conditions for such band crossings in region
III. In this case, since the critical point is crossed again,
$\mu_{\vk}(t)=1$ for all $\vk = \vk_0, \vk'_0$. For the
critical modes, where $p_{\vk_0}=1$, using Eqs.~\eqref{evo1} and
\eqref{eigenreg3}, we find
\begin{eqnarray} && \xi_{\vk_0}(t_{1\vk_0},0)-\xi_{\vk_0}(t_{2\vec
k_0},t_{1\vk_0}) + \xi_{\vk_0}(t_0, t_{2 \vk_0}) = n \pi.
\label{croscon3} \end{eqnarray}
For the non-critical modes, where the instantaneous energy levels do
not cross at $t_{1 \vk}$ or $t_{2 \vk}$, $p_{\vk'_0}=0$ and
we find from Eq.\ \eqref{eigenreg3} that the crossing condition is
given by Eq.\ \eqref{croscon1} with $t_0 \ge t_{2 \vk}$.

Eqs.\ \eqref{croscon1}, \eqref{croscon2} and \eqref{croscon3} constitute
general conditions for phase band crossings for the integrable models
that we study. Although we have obtained them using the
adiabatic-impulse approximation, they are essentially exact since the
adiabatic-impulse approximation becomes exact for modes for which
$\Delta_{\vk_0}=0$. Our results also demonstrate that the
conditions for the phase band crossings at the critical mode (Eqs.\
\eqref{croscon2} and \eqref{croscon3}), where we find an unavoided
crossing of the instantaneous energy levels, is different from those
of the non-critical modes with no instantaneous energy level
crossings. The origin of this difference can be easily traced to the
fact that at each such crossing $g(t)-b_{\vk_0}$ changes sign;
thus the sign of the phase accumulated is reversed in region II
which leads to the difference between Eqs.\ \eqref{croscon2} and
\eqref{croscon3} with Eq.\ \eqref{croscon1}. We also note that the
difference in phase band crossing conditions for the critical modes
that we unravel here is absent for all protocols where the critical
point/region is traversed instantly, {\it i.e.}, at an infinite
rate. Examples of such protocols include periodic arrays of $\de$-function
kicks and square pulses studied earlier
\cite{thakurathi1,thakurathi2,rudner1}; for these protocols $t_{2 \vec
k}=t_{1 \vk}$ for all $\vk$. Consequently Eqs.\ \eqref{croscon2}
and \eqref{croscon3} become identical to Eq.\ \eqref{croscon1}.

\begin{figure}
\includegraphics[width=0.47\linewidth]{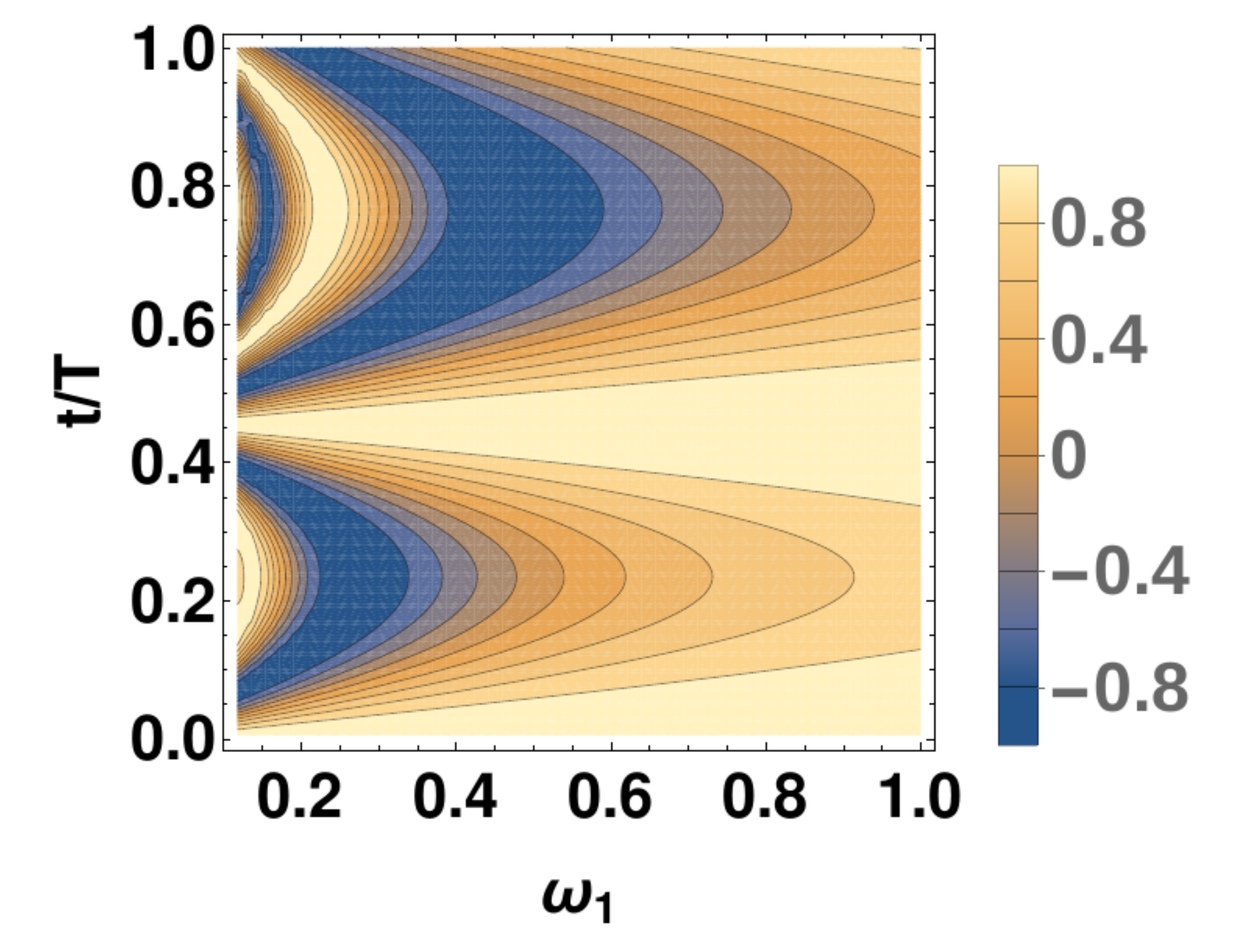}
\includegraphics[height=3.2cm, width=0.47\linewidth]{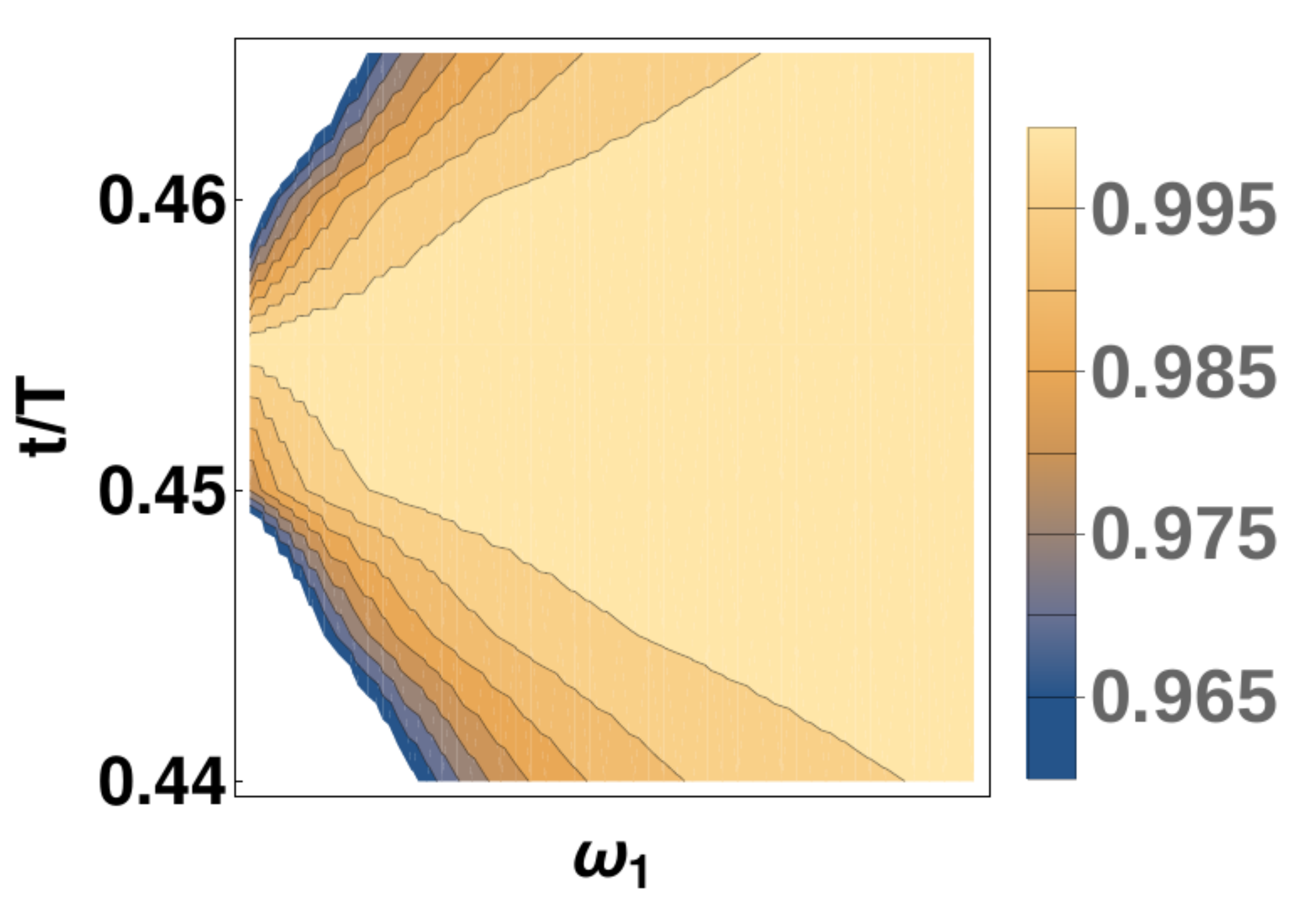}
\caption{Left Panel: A plot of $\cos(\phi_0(t))$ as a function of
$\om_D$ and $t/T$ for the 1D transverse field Ising model for the
drive protocol $h(t)= h_0 + h_1 \cos(\om_D t)$, with $h_0=1.1$, and
$h_1=-1$. Note the extended region around $t/T=0.45$ where
$\phi_0(t)$ vanishes independently of the value of $\om_D$. See text
for details. Right Panel: A closer look at a phase band crossing
corresponding to $n=1$ highlighting the local nature of the crossing
as a function of $t/T$ and $\omega_D$.} \label{fig4} \end{figure}

\begin{figure}
\includegraphics[width=\linewidth]{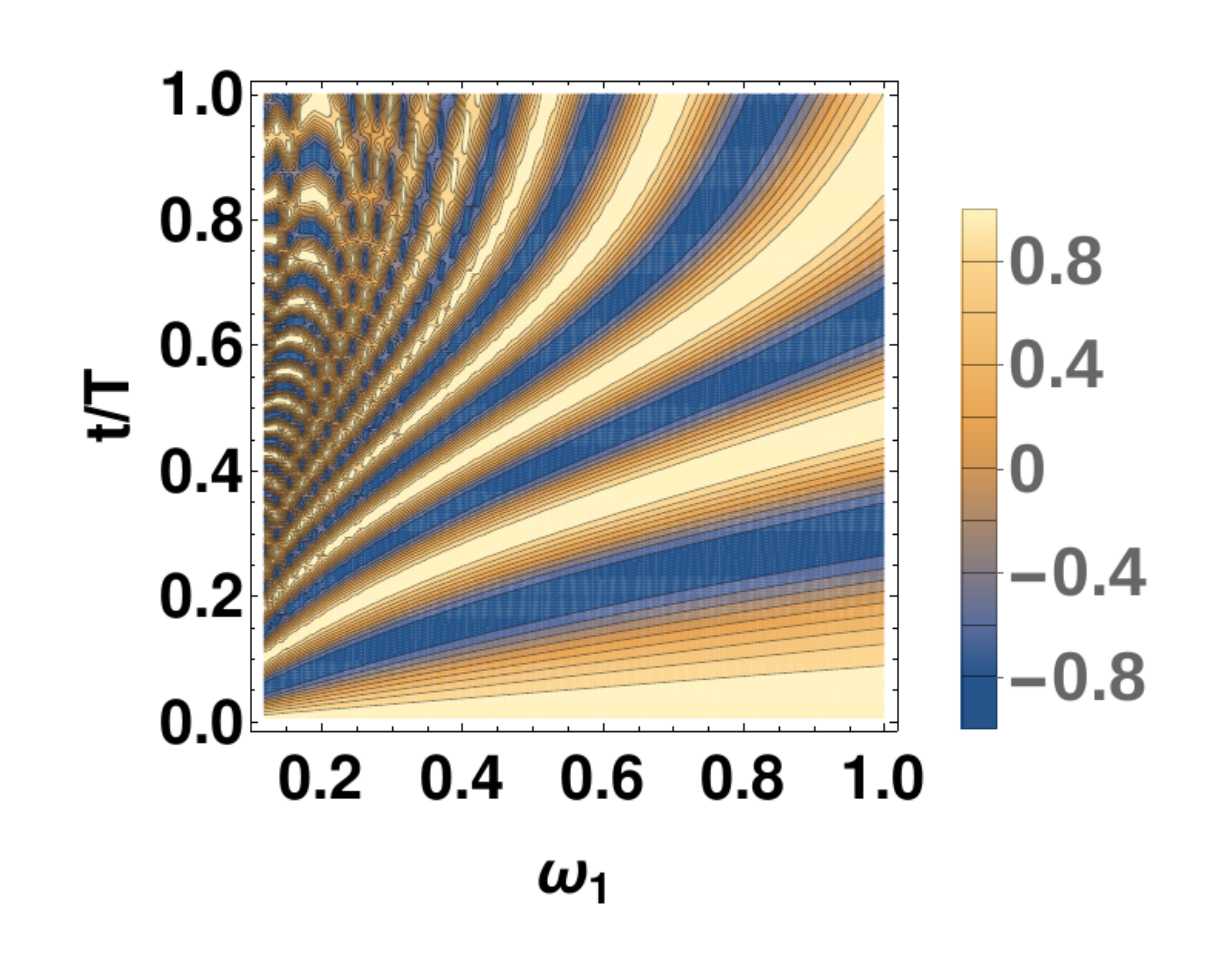}
\caption{A plot of $\cos(\phi_{\pi}(t))$ as a function of $\om_D$
and $t/T$ for the 1D transverse field Ising model for the drive
protocol $h(t)= h_0 + h_1 \cos(\om_D t)$ with $h_0=1.1$, and
$h_1=-1$. Note that in contrast to Fig.\ \ref{fig4}, the phase band crossings
(bright yellow regions) depend on $\om_D$.} \label{fig5} \end{figure}

Having established the crossing conditions for a generic protocol,
we now study specific examples of such crossing for the 1D
transverse field Ising model and the 2D Kitaev model. For the transverse
field Ising model, the crossings occur at $k=0$ or $k=\pi$. For the
latter mode, there is no band crossing and the crossing condition is
given by Eq.\ \eqref{croscon1}, while for the former mode, the bands
cross when $h(t)=1$ and the crossing conditions in regions II and
III are given by Eqs.\ \eqref{croscon2} and \eqref{croscon3}. Using the
protocol $h(t)=h_0 + h_1 \cos(\om_D t)$, we find that the band
crossing conditions at $k= \pi$ in all three regions is given by
\begin{eqnarray} h_1 \sin(x) + (h_0 +1) x &=& n \pi \om_D,
\label{kpicondition} \end{eqnarray}
where $x= \om_D t$ and $n$ is an integer. For the $k=0$ mode, the
conditions in region I, II and III can be obtained from Eqs.\
\eqref{croscon1}, \eqref{croscon2} and \eqref{croscon3}. Noting that
$E_0(t) = |h(t)-1|$ and $h(t)-1$ changes sign at $t_1$ and $t_2$,
we can combine the conditions in Eqs.\ \eqref{croscon1}, \eqref{croscon2},
and \eqref{croscon3} to obtain
\begin{eqnarray} h_1 \sin(x) + (h_0 -1) x &=& n \pi \om_D.
\label{k0condition} \end{eqnarray}
We note that for $n=0$, the crossing conditions imply that
$\om_D t_0$ is constant which indicates that such a crossing spans
over a range of frequencies. This observation is verified from the plot
of the phase bands at $k=0$ as a function of $\om_D$ and $t/T$.
We find that the crossing time $t/T$ corresponding to
$\phi_0(t_0)=0$ which occurs in region II is independent of $
\om_D$; thus $\cos[\phi_0(t)] =1 $ for any $\om_D$ at
$t_0/T=0.45$ as can be seen in Fig.\ \ref{fig4}. A similar plot for
$k=\pi$ (Fig. \ref{fig5}) does not show this behavior; for $k= \pi$
and for our choice of $h_0$ and $h_1$ the crossing always occur at
$n \ne 0$ (Eq.\ \eqref{kpicondition}).

Next, we consider the Kitaev model in $d=2$.~\cite{kit1,nussinov1} This model
consists of spin-1/2's on a honeycomb lattice with nearest-neighbor
interactions described by the Hamiltonian
\bea H &=& \sum_{j+l={\rm even}} \left( J_1 \si_{j+1,l}^x \si_{j,l}^x ~+~ J_2
\si_{j-1,l}^y \si_{j,l}^y \right. \non \\
&& ~~~~~~~~~~~~~~~~\left. + ~J_3 \si_{j,l+1}^z \si_{j,l}^z \right),
\label{kitham} \eea
where $(j,l)$ denotes coordinates of a site on the honeycomb lattice, and
$J_{1,2,3}$ are the couplings between the $x,y,z$ components of neighboring
spins. The unit cell of the system has two sites which we label as $a$ and
$b$. Denoting the location of a
unit cell by $\vn$, a Jordan-Wigner transformation takes us from spin-1/2's to
two Hermitian (Majorana) fermion operators in each unit cell labeled
as $a_\vn$ and $b_\vn$. We can go to momentum space by defining
\bea a_\vn &=& \sqrt{\frac{4}N{}} ~\sum_\vk ~[a_\vk ~e^{i \vk \cdot \vn} ~+~
a_\vk^\da ~e^{-i \vk \cdot \vn}], \non \\
b_\vn &=& \sqrt{\frac{4}N{}} ~\sum_\vk ~[b_\vk ~e^{i \vk \cdot \vn} ~+~
b_\vk^\da ~e^{-i \vk \cdot \vn}], \label{abkn} \eea
where $N$ is the number of sites (the number of unit cells is $N/2$), and
the sum over $\vk$ goes over half the Brillouin zone.
A convenient choice of the Brillouin zone is given by a rhombus whose
vertices lie at $(k_x,k_y)= (\pm 2\pi, 0)$ and $(0,\pm 2\pi /\sqrt{3})$;
half the Brillouin zone is then an equilateral triangle with vertices at
$(0,\pm 2\pi /\sqrt{3})$ and $(2\pi,0)$.

Eq.~\eqref{kitham} leads to a fermionic Hamiltonian of the form
\bea H &=& \sum_\vk ~\left( \begin{array}{cc}
a_\vk^\da & b_\vk^\da \end{array} \right) ~H_\vk ~\left(
\begin{array}{c}
a_\vk \\
b_\vk \end{array} \right), \non \\
H_\vk &=& 2 [J_1 \sin(\vk \cdot \vec{M}_1)-J_2 \sin(\vk \cdot \vec{M}_2)]
\tau^x \non \\
&& + 2 [J_3+J_1 \cos(\vk \cdot \vec{M}_1)+J_2 \cos(\vk \cdot \vec{M}_2)]
\tau^y, \label{hkitaev} \eea
where $\vec{M}_1 = \frac{1}{2} \hat{i} + \frac{\sqrt{3}}{2} \hat{j}$ and
$\vec{M}_2 =\frac{1}{2} \hat{i} - \frac{\sqrt{3}}{2} \hat{j}$ are spanning
vectors which join some neighboring unit cells ($\hat{i}$ and $\hat{j}$
denote unit vectors in the $x$ and $y$ directions, and we have taken the
nearest-neighbor spacing to be equal to $1/\sqrt{3}$), and $\tau^a$ are
Pauli matrices in the $a, ~b$ space.

We now drive $J_3$ is periodically in time as \beq J_3 (t) ~=~ h_0
~+~ h_1 \cos (\om_D t). \label{j3t} \eeq To see what happens to the
phase bands, let us consider the case $J_1 = J_2$ for simplicity. We
then obtain
\bea H_\vk &=& 4 J_1 \cos (k_x/2) \sin (\sqrt{3} k_y/2) \tau^x \non \\
&& + [2J_3 (t) + 4 J_1 \cos (k_x/2) \cos (\sqrt{3} k_y/2)] \tau^y.
\label{hamkit} \eea The form of this is similar to that in
Eq.~\eqref{fermhamden}, except that $\tau^z$ has been replaced by
$\tau^y$; indeed these two Hamiltonians are related to each other by
a global unitary transformation \cite{ks1}. By the arguments given
earlier, we therefore see that phase band crossings will occur at a
momentum $\vk_0$ and time $t_0$ given by
\bea \cos (k_{0x}/2) \sin (\sqrt{3} k_{0y}/2) &=& 0, \label{phasebandkitaev} \\
(h_0 + 2 J_1 \cos (\frac{k_{0x}}{2}) \cos (\frac{\sqrt{3}
k_{0y}}{2}))x + h_1 \sin(x) &=& \frac{n \pi \om_D}{2}, \non \eea
where $x= \om_D t_0$. We now observe that the first equation in
Eq.~\eqref{phasebandkitaev} is satisfied for momenta lying on one of
two lines in half the Brillouin zone; the two lines are given by \\ \\
\noi (i) $k_{0x} = \pi$ and $-\pi/\sqrt{3} \le k_{0y} \le \pi/\sqrt{3}$, \\
\noi (ii) $k_{0y} = 0$ and $0 \le k_{0x} \le 2 \pi$. \\ \\
Correspondingly, the second equation in Eq.~\eqref{phasebandkitaev}
implies that we must have \beq h_0 x ~+~ h_1 \sin (x) ~=~ \frac{n
\pi \om_D}{2} \label{line1} \eeq on line (i), and \beq (h_0 + 2
J_1 \cos (k_{0x}/2))x ~+~ h_1 \sin (x) ~=~ \frac{n \pi \om_D}{2}
\label{line2} \eeq on line (ii). Interestingly,
Eqs.~(\ref{line1}-\ref{line2}) show that the phase band crossing
time $t_0$ is independent of the location of the momentum on line
(i) but depends on the location of the momentum $k_{0x}$ on line
(ii). Analogous conditions for the case $J_1 \ne J_2$ can be
obtained by a similar analysis of Eq.\ \eqref{hamkit};
however analytical expressions similar to Eq. \eqref{phasebandkitaev}
may be difficult to obtain in such cases.

To conclude, in the 2D Kitaev model, phase band crossings occur on
certain lines in momentum space. This is in contrast to 1D models
like the Ising model in a transverse field where phase band
crossings occur only at some discrete momenta, namely, 0 and $\pi$.
This difference constitute a concrete example of the symmetry based
arguments provided in Ref.\ \onlinecite{rudner1}.

\section{Fermionic correlators}
\label{corr1}

In this section, we show that the phase band crossings leave their
imprint on the Fourier transform of some of the off-diagonal
fermionic correlators. To this end, we recall that the
time-dependent Hamiltonians $H$ given by Eqs. \eqref{fermham} and
\eqref{fermhamden}. Here and in the rest of this section, we shall
assume $\Delta_\vk$ to be real for simplicity; however our analysis
may be readily generalized to complex $\Delta_{\vk}$. To obtain
the correlators, we first note that if $(u_{\vk},v_{\vk})^T$
is an eigenvector of $H_\vk$, the eigenstates of $H_{\vk}$ in
second quantized form is given by $(u_{\vk} + v_{\vk} c_{\vec
k}^\da c_{-\vk}^\da ) | vac \ra$. In this state, we find that
$\la c_{\vk}^\da c_{\vk} \ra = \la c_{-\vk}^\da c_{-\vk}
\ra = |v_{\vk}|^2$ and $\la c_{\vk}^\da c_{-\vk}^\da \ra =
u_{\vk} v_{\vk}^*$. In what follows, we rewrite the
expressions for these correlators in a different way so as to point
out their connections to the phase bands. To do this, let us denote
the eigenvectors of $U_{\vk}(t)$ corresponding to the eigenvalues
$\lambda_{\pm \vk}(t) = \exp[i \phi_{\pm \vk}(t)] = \exp[\pm i
\phi_{\vk}(t)]$ as
\begin{eqnarray}
|\chi_{\pm \vk}(t)\rangle = \left( \begin{array}{c} \mu_{\pm \vec
k}(t)
\\ \nu_{\pm \vk}(t) \end{array} \right). \label{eigenvecU}
\end{eqnarray}
Note that $|\chi_{\pm \vk}(t) \rangle$ forms a complete basis.
Using these eigenvectors and eigenvalues, we can now generic
expressions for both the off-diagonal and the diagonal correlators in
terms of these eigenvectors and eigenvalues as
\begin{eqnarray}
C_{\vk}(t) &=& \langle \psi_{\vk}(0)|U^{\dagger}_{\vk}(t)
c_{\vk}^{\dagger} c_{\vk}U_{\vk}(t) |\psi_{\vk}(0) \rangle \non \\
&=& \sum_{a,b= \pm} (u^0_{\vk} \mu_{a \vk}(t) + v_{\vk}^0
\nu_{a\vk}(t)) \nu^{\ast}_{a \vk}(t) \nu_{b\vk}(t) \non \\
&& \times e^{-i(\phi_{a \vk}(t)- \phi_{b \vk}(t))} (u^0_{\vec
k} \mu_{b \vk}(t) + v_{\vk}^0 \nu_{b\vk}(t)), \non \\
F_{\vk}(t) &=& \langle \psi_{\vk}(0)|U^{\dagger}_{\vk}(t)
c^{\dagger}_{-\vk} c^{\dagger}_{\vk} U_{\vk}(t) |\psi_{\vec
k}(0) \rangle \non \\
&=& \sum_{a,b= \pm} (u^0_{\vk} \mu_{a \vk}(t) + v_{\vk}^0
\nu_{a\vk}(t)) \mu^{\ast}_{a \vk}(t) \nu_{b\vk}(t) \non \\
&& \times e^{-i(\phi_{a \vk}(t)- \phi_{b \vk}(t))} (u^0_{\vec
k} \mu_{b \vk}(t) + v_{\vk}^0 \nu_{b\vk}(t)). \label{corexp0}
\end{eqnarray}
Thus we find that the phase bands contribute to the terms in the correlators
for $a \ne b$. For $\vk= \vk_0$, for which $|\chi_{\pm \vk}(t)
\rangle$ are eigenstates of $\tau_z$ at all times, time-dependent
contributions from the phase bands only appear in $F_{\vk}(t)$, since
$C_{\vk}(t)$ receives contribution only from the $a=b$ terms
which are time-independent. Finally, we note that for $d=1$ where the phase
bands cross at $k=0, \pi$, where $c_{k_0}^{\dagger} c_{-k_0}^{\dagger}=0$ due
to Pauli exclusion; however these correlators are non-zero for $d>1$ where
such crossing may occur at $\vk_0 \ne 0, \pi$.

For $\vk = \vk_0$, the phase bands are given by $\phi_{\pm \vk_0}(t) = \pm
\int_0 ^t dt' (g(t') -b_{\vk_0})$, and the corresponding eigenvectors are
$|\chi_{+ \vec k_0}(t)\rangle = (1,0)^T$ and $|\chi_{- \vk_0}(t)\rangle=
(0,1)^T$. Substituting these in Eq.\ \eqref{corexp0}, we obtain
$C_{\vk_0}(t) = |v^0_{\vk_0}|^2$ (which is independent of time) and
\begin{eqnarray}
F_{\vk_0}(t) = u_{\vk_0}^0 v_{\vk_0}^{0 \ast} \exp [-2 i \int_0^t dt' (g(t')
- b_{\vk_0})]. \label{corexp1} \end{eqnarray}
We note that if the phase bands cross at $t=t_0$, we have $F_{\vec
k_0}(t_0)= F_{\vk_0}(0) = u_{\vk_0}^0 v_{\vk_0}^0$. In what
follows, we study the Fourier transform of these correlators given by
\begin{eqnarray}
F_{\vk_0}(\om_0) = \int_0^T dt e^{i \om_0 t} F_{\vk_0} (t),
\label{correxp2} \end{eqnarray}
and show that the phase band crossings leave their imprint on the Fourier
transform.

For the sake of concreteness, we apply these ideas to the Kitaev
model with the Hamiltonian given in Eq.~\eqref{hamkit}. As remarked
earlier, the structure of this is similar to \eqref{fermham} except
that $\tau^y$ is replaced by $\tau^z$, and the Kitaev model has an
off-diagonal couplings like $a_\vk^\da b_\vk$ (which conserve
fermion number) instead of superconducting pairing terms like
$c_\vk^\da c_{-\vk}^\da$. We can take care of this difference by
transforming to the basis of $\tau^y$ given by
\beq c_\vk ~=~ \frac{1}{\sqrt{2}} ~(a_\vk + i b_\vk), ~~~~d_\vk ~=~
\frac{1}{\sqrt{2}} ~(a_\vk - i b_\vk). \eeq
At time $t=0$, we take the wave function to be $(u_{\vk}^0,v_{\vk}^0)^T$
which denotes the state $(u_{\vk}^0 c_\vk^\da + v_{\vk}^0
d_\vk^\da) | vac \ra$ in second quantized notation. We can then
calculate the time-dependent fermionic correlators in the Kitaev model
in the same way as in the previous paragraph.

\begin{figure}
\includegraphics[width=0.47\linewidth]{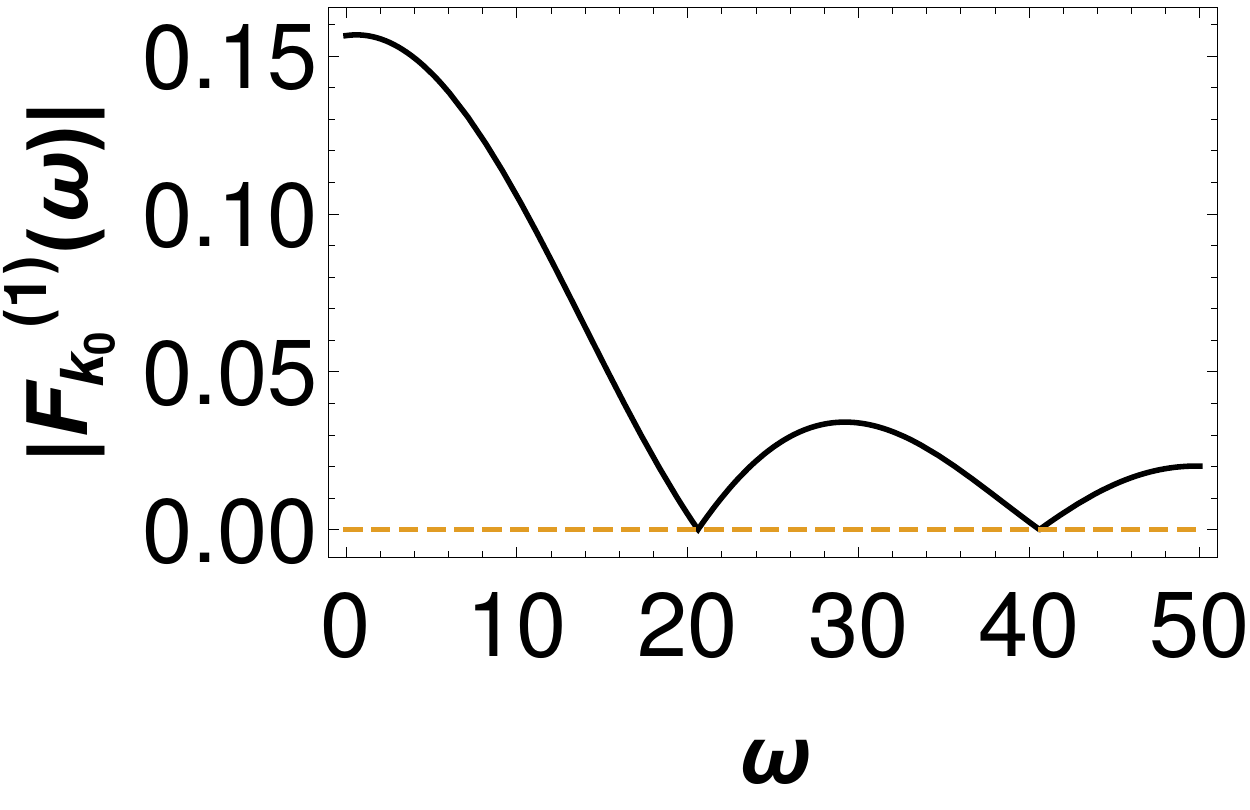}
\includegraphics[width=0.47\linewidth]{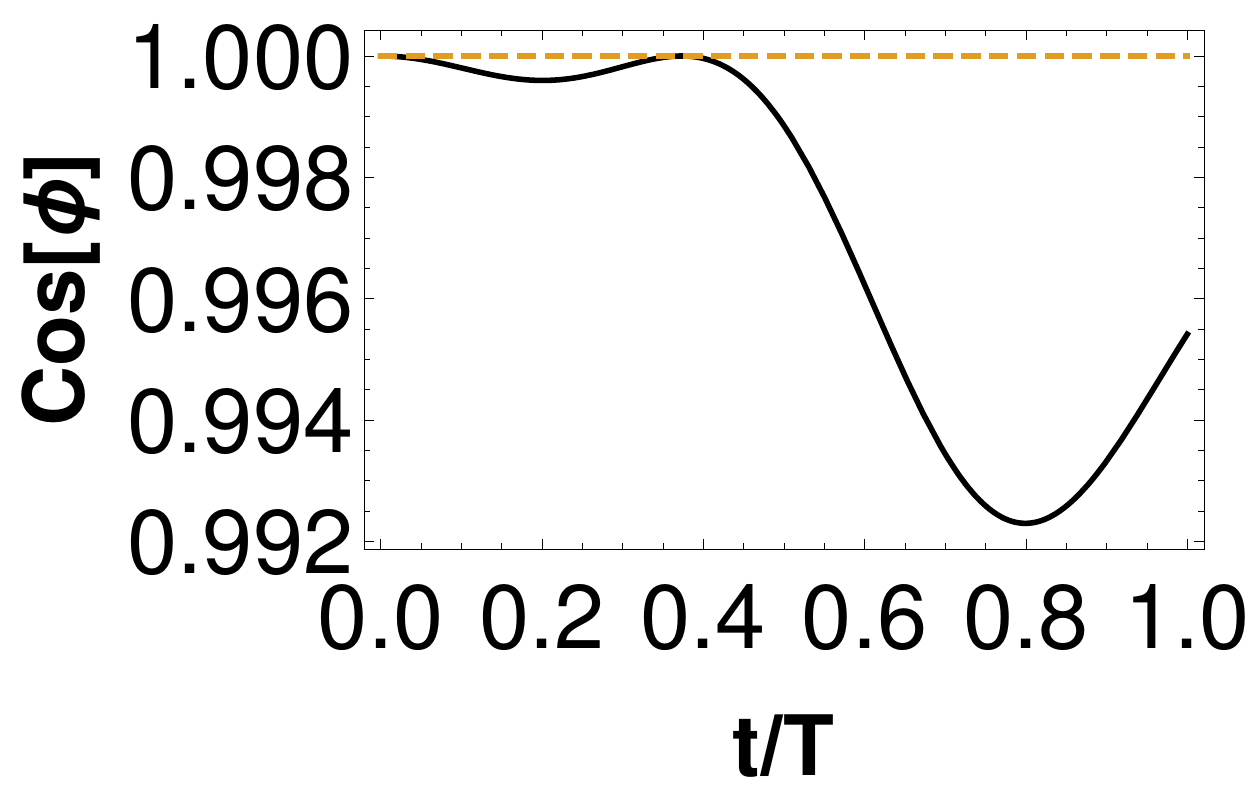}
\caption{Left Panel: Plot of $|F^{(1)}_{\bf k_0}(\omega_0)|$ vs
$\omega_0$ for the Kitaev model showing maxima and minima of the
off-diagonal correlation function with $(k_{x0}, k_{y0})=
(\pi/9,0)$. The protocol used is $J_3(t)= h_0 + h_1 \cos(\omega_D
t)$ with $h_0=h_1=0.5$, and $\omega_D=20$. Here we have chosen
$u_0=v_0=1/\sqrt{2}$, $J_1=J_2$, and all energies are scaled in
units of $J_1$. The right panel shows that phase bands cross at
$t_0=0.4T$; we find that the maxima and minima obtained coincides
with those predicted from Eqs.\ \eqref{maxcond} with $p=0$ (maxima)
and Eq.\ \eqref{mincond} with $p=0$ and $m=1,2$ (minima). The dotted
lines are guides to the eye.} \label{fig6} \end{figure}

For the drive protocol given in Eq.~\eqref{j3t}, we have seen that
phase band crossings can occur when the term proportional to $\tau^x$ in
Eq.~\eqref{hamkit} vanishes, and these happens on one of two lines in
momentum space. For definiteness, let us consider a point $\vk_0$ on
the second line, with $k_{0y} = 0$ and $0 \le k_{0x} \le 2 \pi$.
Using Eq.~\eqref{corexp0} and \eqref{corexp1}, we then find that
$\la c_{\vk_0}^\da c_{\vk_0} \ra_t$ and $\la d_{\vk_0}^\da d_{\vk_0} \ra_t$
are independent of time. In contrast the off-diagonal correlator is given by
$F_{\vk}(t) = \langle c_{\vk}^{\dagger} d_{\vk}\rangle_t$ and reads
\bea F^{(1)}_{\vk_0}(t) &=& u_0^* v_0 \exp [- i\om_{\vk_0} t + i
(4 h_1 /\om_D) \sin (\om_D t)], \non \\
\om_{\vk_0} &=& - 4 h_0 - 8 J_1 \cos (k_{0x}/2) \label{fkt1} \eea
The Fourier transform of \eqref{fkt1} for one time period $T$ is
given by \beq F^{(1)}_{\vk_0}(\om_0) ~=~ - i u_0^* v_0
\sum_{n=-\infty}^\infty J_n \left( \frac{4h_1}{\om_D} \right)
~\frac{e^{i (\om - \om_{\vk_0} + n \om_D) T} ~-~ 1}{\om -
\om_{\vk_0} + n \om_D}, \label{fkom1} \eeq where we have used the
identity \cite{abram} \beq e^{i z \sin \theta} ~=~
\sum_{n=-\infty}^\infty ~J_n (z) ~e^{i n \theta}. \eeq If
$4h_1/\om_D \ll 1$, the $n=0$ term will dominate in the sum in
Eq.~\eqref{fkom1}. We then find that as a function of $\om$, the
magnitude of $F^{(1)}_{\vk_0} (\om)$ has a maximum at $\om_0 \simeq
\om_{\vk_0}$ where $|F^{(1)}_{\vk_0} (\om_0)| \simeq |u_0^* v_0 T J_0
(4h_1/\om_D)|$, and minima at $\om_{0m} \simeq \om_{\vk_0} + m \om_D$
(with $m$ being a non-zero integer) where $F^{(1)}_{\vk_0} (\om_{0m}) = 0$.
Since the phase band crossings occur at times $t_0$ given by
$\om_{\vk_0} t_0 - (4 h_1 /\om_D) \sin (\om_D t_0) = 2 p \pi$ (where
$p \in Z$), we therefore obtain that $F^{(1)}_{\vk_0}(\om_0)$ will be
maximum if
\begin{eqnarray} \om_0 t_0 &=& (4 h_1/\om_D) \sin[\om_D t_0] + 2 \pi p,
\label{maxcond} \end{eqnarray}
and display minima at
\begin{eqnarray} \om_{0m} t_0 &=& (4 h_1/\om_D) \sin[\om_D t_0] +m
\om_D t_0 + 2 \pi p \label{mincond} \end{eqnarray} for non-zero
integer $m$. Thus the maxima-minima pattern of $F^{(1)}_{\vec
k_0}(\om_0)$ contains information about the phase band crossing
time. The precise relation between $t_0$, $\om_D$ and $\om_0$ which
leads to these maxima and minima depends on the drive protocol used.
This is demonstrated in Fig.\ \ref{fig6} where $|F_{\vec
k_0}^{(1)}(\omega)|$ is plotted as a function of the frequency
$\omega$ in the left panel; the right panel shows the position of
$t_0$ as obtained from Eqs.\ \eqref{line1} and \eqref{line2} for the
specific drive parameters used. We note that the maxima and the
minima of $|F_{\vec k_0}^{(1)}(\omega)|$ occurs at frequencies which
are given by Eq.\ \eqref{maxcond} and \eqref{mincond} with $t_0=0.4
T$.

To elucidate the protocol dependence stated above, we consider a
drive protocol consisting of periodic $\de$-function
kicks~\cite{thakurathi2}, \beq J_3 (t) ~=~ h_0 ~+~ h_1
~\sum_{n=-\infty}^\infty ~\de (t - nT), \eeq for which the phase
band crossings occur at $\om_{\vk_0} t_0 - 4 h_1 = 2 n \pi$, where
$\om_{\vk_0}$ is given in Eq.~\eqref{fkt1}. A calculation similar to
the one outlined above yields \beq F^{(2)}_{\vk_0}(\om) ~=~ - i
u_0^* v_0 ~e^{i 4 h_1} ~\frac{e^{i (\om - \om_{\vk_0}) T} ~-~ 1}{\om
- \om_{\vk_0}}. \label{fkom2} \eeq The magnitude of
$F^{(2)}_{\vk_0}(\om_0)$ has a maximum at $\om_0 = \om_{\vk_0}$
where $|F^{(2)}_{vk_0}(\om_0)| = |u_0^* v_0 T|$, and minima at
$\om_{0m} = \om_{\vk_0} + m \om_D$ (with $m \ne 0$) where
$F^{(2)}_{\vk_0}(\om_{0m}) = 0$. This leads to the relations
\begin{eqnarray} \om_0 t_0 = 4 h_1 + 2 \pi n \label{maxkick}\end{eqnarray}
for maxima, and
\begin{eqnarray} \om_{0m} t_0 = 4 h_1 + 2 \pi n + m \om_D \label{minkick}\end{eqnarray}
for minima of $F^{(2)}_{\vk_0}(\om_0)$.

\begin{figure}
\includegraphics[width=0.47\linewidth]{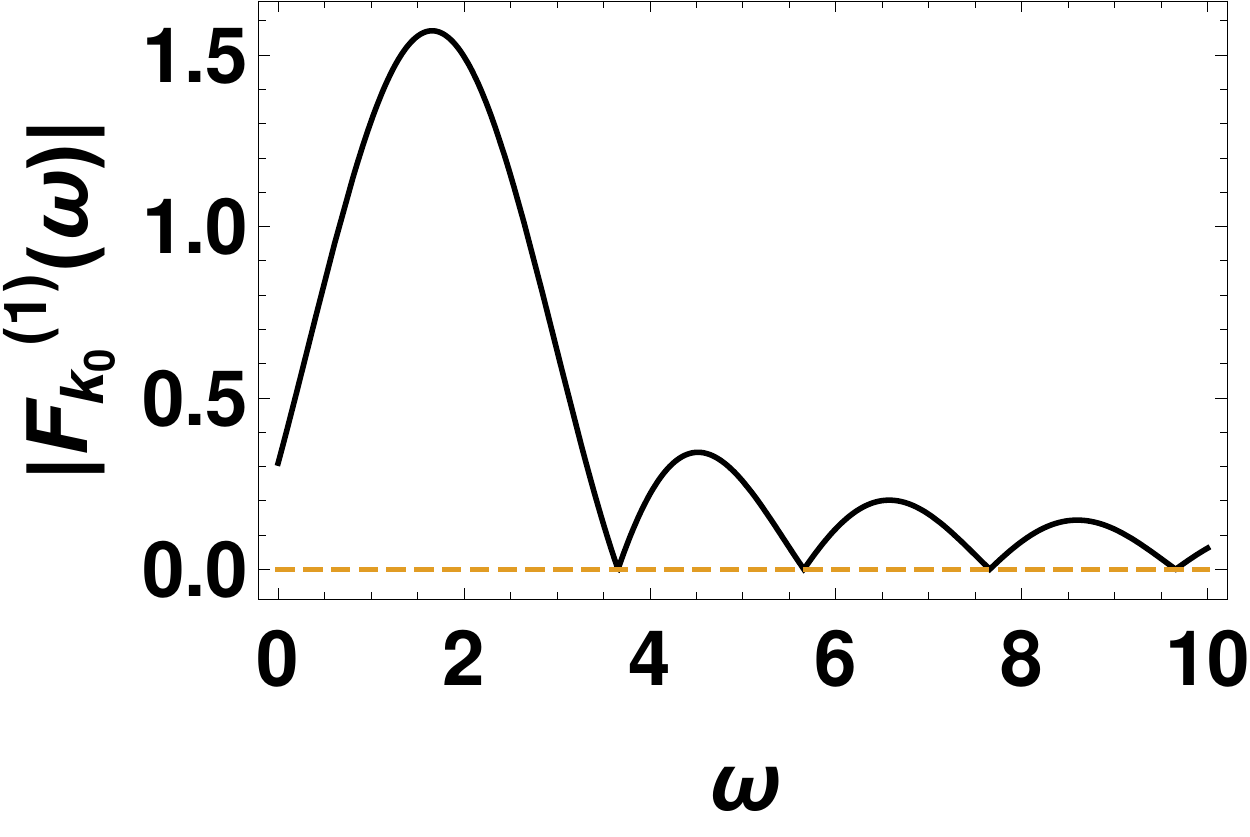}
\includegraphics[width=0.47\linewidth]{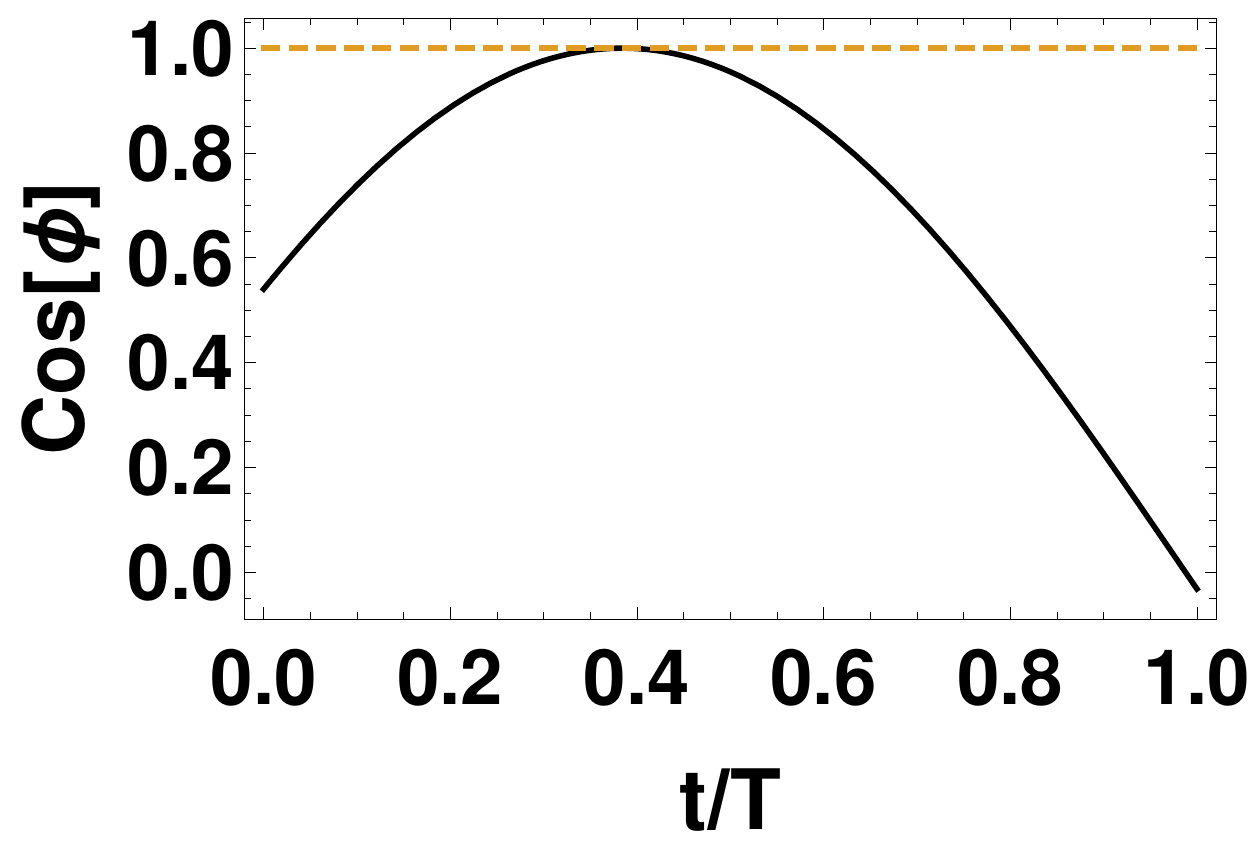}
\caption{Left Panel: Plot of $|F^{(2)}_{\bf k_0}(\omega_0)|$ vs
$\omega_0$ for the Kitaev model showing maxima and minima of the
off-diagonal correlation function for a delta function protocol with
periodic kicks with $(k_{x0}, k_{y0})= (\pi/2,0)$, $h_0=1$ and
$h_1=0.5$. All other parameters are the same as in Fig.\ \ref{fig6}.
The right panel shows that $t_0=0.41 T$; we find that the maxima and
minima obtained coincides with those predicted from Eqs.\
\eqref{maxkick} with $n=0$ (maxima) and Eq.\ \eqref{minkick} with
$n=0$ and $m=1,2,3$ (minima). The dotted lines are guides to the
eye.} \label{fig7} \end{figure}

A plot of $|F_{\vec k_0}^{(2)}(\omega)|$ is plotted as a function of
the frequency $\omega$ in the left panel; the right panel shows the
position of $t_0$ for the specific drive parameters used. We note
that the maxima and the minima of $|F_{\vec k_0}^{(1)}(\omega)|$
occurs at frequencies which are given by Eq.\ \eqref{maxkick} and
\eqref{minkick} with $t_0=0.41 T$.

Before ending this section, we note that the fermionic correlators
in the Kitaev model can be related to the correlators of the spins
appearing in the Hamiltonian in Eq.~\eqref{kitham}.\cite{ks1} For
two neighboring sites given by $b$ located at $\vn$ and $a$ located
at $\vn + \vcr$ (where $\vcr$ can take three possible values given
by $(0,0)$, $-{\vec M}_1$ and ${\vec M}_2$), we can use
Eq.~\eqref{abkn} to relate the fermionic correlators in real and
momentum space, \beq \la i b_\vn a_{\vn + \vcr} \ra_t ~=~
\frac{i4}{N} ~\sum_\vk ~\la b_\vk^\da a_\vk e^{i \vk \cdot \vcr} ~-~
a_\vk^\da b_\vk e^{-i \vk \cdot \vcr} \ra_t, \eeq where we have used
the relations $\la b_\vk a_{\vk'} \ra_t = 0$ for all $\vk$, $\vk'$,
and $\la b_\vk^\da a_{\vk'} \ra_t = 0$ if $\vk \ne \vk'$. Rewriting
$a_\vk$ and $b_\vk$ in terms of $c_\vk$ and $d_\vk$, we find that
the correlator is given by \bea \la i b_\vn a_{\vn + \vcr} \ra_t &=&
\frac{4}{N} ~\sum_\vk ~ [ \cos (\vk \cdot \vcr) (\la d_\vk^\da d_\vk
\ra_t - \la c_\vk^\da c_\vk \ra_t)
\non \\
&& ~~~~ -~i \sin (\vk \cdot \vcr) (\la c_\vk^\da d_\vk \ra_t - \la
d_\vk^\da c_\vk \ra_t) ]. \label{abn} \eea We thus see that the
terms proportional to $\sin (\vk \cdot \vcr)$ are related to the
off-diagonal fermion correlators. Next, we note that for $\vcr =
(0,0)$, $-{\vec M}_1$ and ${\vec M}_2$, $i b_\vn a_{\vn + \vcr}$ are
given by $\si^z_{b,\vn} \si^z_{a,\vn}$, $\si^y_{b,\vn} \si^y_{a,\vn
- {\vec M}_1}$ and $\si^x_{b,\vn} \si^x_{a,\vn + {\vec M}_2}$
respectively~\cite{ks1}. Hence in the last two cases, where $\vcr =
-{\vec M}_1$ or ${\vec M}_2$, the nearest-neighbor spin correlators
are related to the off-diagonal fermion correlators through
Eq.~\eqref{abn}. (If $b_\vn$ and $a_{\vn + \vcr}$ are not on
nearest-neighbor sites, the relation between $i b_\vn a_{\vn +
\vcr}$ and spin correlators is more complicated. Namely, $i b_\vn
a_{\vn + \vcr}$ is given by a product of $\si^x$ or $\si^y$ at $\vn$
and $\vn + \vcr$ multiplied by a Jordan-Wigner string of $\si^z$'s
running between those two sites).

To summarize, at the momenta $\vk_0$ where phase band crossings can
occur, we find that the Fourier transform of the off-diagonal
correlator, $F_{\vk_0}(\om_0)$ has maxima and minima at some
particular frequencies; these frequencies are related to
$\om_{\vk_0}$ (which is a function of $\vk_0$) by integer multiples
of the drive frequency $\om_D$. The maxima and minima of these correlators
provide us with a relation between $\om_0$, the phase
band crossing time $t_0$, and the drive frequency $\om_D$ whose
precise form depends on the drive protocol used. We note that the
standard signature of the phase band crossings shows up in the form of
localized subgap states at the ends of a finite sample
\cite{rudner1}; however, such a signature cannot identify the
crossing time $t_0$ which can be done by tracking maxima and minima
of $F_{\vk_0} (\om_0)$.

\section{Discussion}
\label{diss}

In this work, we have provided an analytic expression for the
phase bands for a class of periodically driven integrable models for
arbitrary drive protocols within the adiabatic-impulse approximation.
Using this expression, and other more generic arguments, we have
outlined the conditions for phase band crossings in these models for
arbitrary drive protocols. As we have argued in this work, although the
expressions for the phase bands are derived within the
adiabatic-impulse approximation, the crossing conditions derived are
exact for two reasons. First, such conditions can be derived from
intuitive arguments which do not depend on the approximations used
and second, the momenta at which these crossings occur are the ones
in which the adiabatic-impulse approximation becomes exact. We also
show that for a class of these crossings, the time of crossing
$t_0/T$ is independent of the drive frequency $\om_D$, and we provide
an analytical explanation of this phenomenon. We also point out that
the crossing conditions for the critical modes, where the
instantaneous energy levels of the Hamiltonian undergo an unavoided
level crossing, are different compared to those for the non-critical
modes where no such level crossings occur. Finally, we point out
that the off-diagonal fermionic correlators carry a signature of
the phase band crossings, and we provide analytical relations between
the frequencies $\om_0$ (at which $F_{\vk_0}(\om_0)$ either
shows a maxima or vanishes), the phase band crossing time $t_0$, and
the drive frequency $\om_D$.

Our results regarding the phase band crossing conditions have some
implications which we briefly discuss. First, although our
results are derived for $N=2$ phase bands, they may provide an
insight into generic crossing conditions for systems with $N>2$. To
see this, let us consider a situation where there are $N$ phase
bands for any given quasi-momentum $\vk$ in a system. Let us
consider a phase band crossing corresponding to a zone-edge
singularity between the top ($N^{\rm th}$) and the bottom ($1^{\rm
st}$) bands. The dynamics of these bands can always be described by
an effective $2 \times 2$ matrix Hamiltonian $H_{\vk}^{\rm eff}$
which can be obtained, in principle, by integrating out all other
degrees of freedom of the system Hamiltonian. Then the effective
evolution operator describing the crossing of these two bands may
always be written, sufficiently near the band crossing point, as $
U_{\vk}^{\rm eff}(t) = {\mathcal T}_t \exp[- (i/\hbar) \int_0^t dt'
H^{\rm eff}_{\vk} (t')]$. The most general form of such an
effective Hamiltonian is given by
\begin{eqnarray} H_{\vk}^{\rm eff} = \sum_{i=1,3} g_{i \vk}(t) \tau_i,
\end{eqnarray}
where $g_{i \vk}(t)$ are parameter functions which depend on $\vk$
and $t$ and whose precise form depends on the details of the system
Hamiltonian and the drive protocol. In terms of these $g_{i \vk}$'s,
the condition for such band crossings, as can be inferred from our
results for integrable models of a similar form of $H$, is that at
least two of the three $g_{i \vk}$ (which we may choose to label as
$g_{1 \vec k}(t)$ and $g_{2 \vk}(t)$ without loss of generality) are
zero for $\vk= \vk_0$ and for appropriate choice of Hamiltonian
parameters. The crossing time $t_0$ is then determined from the
third generator using
\begin{eqnarray} \int_0^{t_0} dt' g_{3 \vk_0}(t') = 0. \end{eqnarray}
Our results for integrable models show that these conditions need
to be satisfied for any $N$ for the system to have a band crossing
corresponding to a zone-edge singularity.

The second implication of our results constitutes the relation of
the symmetry classes of the underlying Hamiltonian to the condition for
phase band crossings. Our results indicate that Hamiltonians
belonging to the same symmetry class \cite{ryu1} and driven by
identical protocols may have different behaviors of the phase bands.
This becomes evident by considering the class CI which contains
models of time-reversal and SU(2) symmetric superconductors that
include both $d$- and $s$-wave pairing symmetries \cite{ryu1}. The
Hamiltonians of such superconductors are given by Eqs.\
\eqref{fermham} and \eqref{fermhamden} where $g = -\mu_0$ and
$b_{\vk} = \epsilon_{\vec k}$, and $\mu_0$ and $\epsilon_{\vk}$ are
the chemical potential and energy dispersion of the fermions. For
$d$-wave superconductors $\Delta_{\vk} = \Delta_0 (k_x^2
-k_y^2)/k_F^2$, while for $s-$wave $\Delta_{\vk}= \Delta_0$. Now
consider periodically driving such a system by changing the chemical
potential: $\mu_0 \equiv \mu_0(t)$. The phase bands corresponding to
$U_{\vk}(t)$ may cross at momenta given by four isolated points on
the Fermi surface $\vk_0 = (k_{0x}, k_{0y})= k_F(\pm 1, \pm
1)/\sqrt{2}$ for $d$-wave superconductors, while they will never
cross for $s-$wave superconductors. Thus the dynamics of these two
models will be very different even if they belong to the same
symmetry class. Our results seem to indicate that for similar
dynamics a necessary condition is that the set of zeroes of two of
the parameter functions $f_{i\vec k}(t)$ or $g_{i\vec k}(t)$ defined
earlier must be identical; for example, if the parameter functions
never vanish so that the set of zeroes is a null set, there will be
no phase band crossing for any drive protocol.

In conclusion, we have presented analytic expressions for the phase bands
for a class of integrable models driven by arbitrary periodic
protocols within the adiabatic-impulse approximation. Using these
expression and other more general arguments, we have listed the
conditions for phase band crossings for such models. We have also
shown that such phase band crossings leave their mark on the Fourier
transform of the off-diagonal fermionic correlators; the positions
of the zeroes and maxima of such correlators provide information about
the band crossing time $t_0$. Finally we have discussed the relevance of
the derived band crossing conditions in the context of generic
models with $N>2$ phase bands, and we have discussed the role of
symmetry in such band crossings.

\vspace{.5cm}

\centerline{\bf Acknowledgments}

\vspace{.5cm}

D.S. thanks DST, India for Project No. SR/S2/JCB-44/2010 for financial support.

\end{document}